\documentclass[a4paper,11pt]{article}
\pdfoutput=1 
\usepackage{jheppub} 
\usepackage[T1]{fontenc} 
\usepackage{physics}
\usepackage{float}

\def\p{{\partial}}      \def\l{{\lambda}}       \def\d{{\delta}}
\def\cL{{\cal L}}              
\def\cM{{\cal M}}              \def\cE{{\cal E}}
       \def\cS{{\cal S}}		\def\cJ{{\cal J}}
\def\cF{{\cal F}}              
              
\def\cV{{\cal V}}

\title{\boldmath Stability of generalized Einstein-Maxwell-scalar black holes}
\author[a]{Radouane Gannouji,}
\author[b,a]{Yolbeiker Rodríguez Baez}

\affiliation[a]{Instituto de Física, Pontificia Universidad Católica de Valparaíso, Av. Brasil 2950, Valparaíso, Chile}
\affiliation[b]{Universidad Técnica Federico Santa María, Av. España 1680, Valparaíso, Chile}

\emailAdd{radouane.gannouji@pucv.cl}
\emailAdd{yolbeiker.rodriguez@gmail.com }

\abstract{We study the stability of static black holes in generalized Einstein-Maxwell-scalar theories. We derive the master equations for the odd and even parity perturbations. The sufficient and necessary conditions for the stability of black holes under odd-parity perturbations are derived. We show that these conditions are usually not similar to energy conditions even in the simplest case of a minimally coupled scalar field. We obtain the necessary conditions for the stability of even-parity perturbations. We also derived the speed of propagation of the five degrees of freedom and obtained the class of theories for which all degrees of freedom propagate at the speed of light. Finally, we have applied our results to various black holes in nonlinear electrodynamics, scalar-tensor theories and Einstein-Maxwell-dilaton theory. For the latter, we have also calculated the quasinormal modes.}

\begin{document} 
\maketitle
\flushbottom
\section{Introduction}

Stability is one of the most important measures in physics, telling us how likely a solution could exist for a long period. It discriminates cosmological solutions from transient situations. Electrovacuum black holes (BHs) in general relativity (GR) are stable. Indeed after the seminal work by Regge and Wheeler \cite{Regge57}, where they proved partially the stability of the Schwarzschild black hole, Zerilli \cite{Zerilli70} provided the stability equations for the even-parity perturbations. Later, following the same approach, Moncrief \cite{Moncrief74-a,Moncrief74-b} derived the stability of the Reissner-Nordstrom black hole. But because of the impossibility to separate variables in the rotating case, Teukolsky \cite{Teukolsky:1972my,Teukolsky:1973ha} developed a new approach which helped to study the linear stability of the Kerr black hole. Even if the Newman-Penrose formalism used by Teukolsky turned out to be extremely difficult to implement for other rotating black holes, such as the Kerr-Newman, the metric approach to non-rotating solution has been widely used in various beyond GR models, see e.g. \cite{DeFelice:2011ka,Takahashi:2010ye,Ganguly:2017ort}. We see that stability is an essential analysis of any solution. On the other hand, thermodynamic stability is also widely studied. It is important to mention that (mechanical) stability should be a condition before any thermodynamic analysis. Indeed, if the black hole is mechanically unstable, any Hawking radiation would destabilize it and therefore render any thermodynamic analysis impossible. Finally we should also mention the interesting aspects of instabilities in nature which provide for example phase transitions. In this direction, any unstable black hole in an asymptotically AdS spacetime could be associated via the AdS/CFT to a phase transition.

From a more astrophysical aspect, BHs turn out to be extremely useful to study models beyond GR, in a strong gravitational regime. As we know, BHs are extremely simple objects. They are described in the vacuum by mainly two parameters, the mass and the angular momentum (the electric charge being usually negligible). These parameters, which describe entirely the BH, can be measured, in particular via the frequency emitted during the last moments of BH mergers, the so-called quasinormal modes (QNMs). Any hairy BH could be discriminated against through these modes. QNMs are easily computed if their equation is known. The analysis performed in this paper will provide these equations. 

As we mentioned, in this paper, we address the problem of black hole stability for a generic class of theories beyond GR. Modifying Einstein's theory of gravity is an easy game but building a new physical theory turns out to be complicated. Many modified gravity theories turn out to be highly constrained by the data, from the Brans-Dicke model to massive gravity à la dRGT. We should follow a simpler road. Could we define theories with sufficient generality? For that, some of the simple elements that we will consider are 
\begin{enumerate}
    \item Ostrogradsky instability free theory, and therefore we will assume second order differential equations for the fields\footnote{Even if higher order differential equation for a scalar field turn out to be possible, see beyond Horndeski \cite{Gleyzes:2014dya,Gleyzes:2014qga}, these theories haven't been very successful in describing the Universe. More generically, degenerate theories \cite{Langlois:2015cwa} can be constructed which do not propagate more degrees of freedom even if higher derivatives are present, by considering an appropriate coupling to the metric which satisfies the degenerate condition.}
    
    \item Well-posed Cauchy problem. Interestingly, this condition reduces drastically the form of the models in presence of a scalar field. Indeed, only K-essence non-minimally coupled survives to these restrictions \cite{Papallo:2017ddx,Papallo:2017qvl,Kovacs:2020ywu}.
\end{enumerate}
Of course, beyond GR theories can't be reduced to only the presence of a scalar field. For example, the dimensional reduction or Kaluza Klein compactification decomposes the higher dimensional metric into at least a vector field along with the scalar field. It is therefore natural to consider a generic model with these two additional degrees of freedom, keeping in mind that K-essence types of theories are strongly hyperbolic. The models studied in this paper can be regarded as a non-linear extension of the hairy black hole studied by Gibbons and Maeda \cite{Gibbons88} in the context of Kaluza-Klein theories, but also rediscovered by Garfinkle, Horowitz and Strominger \cite{Garfinkle91} in the context of string theory. 

In this work, we therefore investigate the linear stability of generalized Einstein-Maxwell-scalar black holes assuming that they are static and without magnetic charge. We will derive some generic results and apply them to specific models. This paper is organized as follows. In section 2, we describe the model and the background equations. In section 3, we summarize the black hole perturbation formalism that we will apply in sections 4 for the odd-parity perturbations and section 5 for the even-parity perturbations. Finally, we will apply our results to various known solutions in the literature, before final comments and conclusions.

To facilitate the use of our results, a {\scshape Mathematica}\textsuperscript{\textregistered} notebook is available online \cite{mathematica}.

\section{Background equations of motion}

We are interested in generalized Einstein-Maxwell-dilaton theories described by the following action
\begin{equation}
    S = \int {\rm d}^4 x \sqrt{-g} \qty[f_1(\phi) \frac{R}{2} + f_2(\phi, X, F)] \, ,
    \label{action}
\end{equation} 
where $R$ is the Ricci scalar, $f_1(\phi)$ is a function of the scalar field $\phi$, representing the non-minimal coupling to gravity, and $f_2(\phi,X,F)$ is a function of the scalar field, its kinetic energy $X = -\frac12 \nabla_\mu\phi \nabla^\mu \phi$ and $F = -\frac14 F_{\mu\nu} F^{\mu\nu}$ where the field strength is defined as $F_{\mu\nu}=\partial_\mu A_\nu - \partial_\nu A_\mu$.

We will study the linear stability of static spherically symmetric spacetimes. 
For that, we will consider a background spacetime described by the following line element\footnote{We will use the notation $\bar{X}$ to refer to quantities evaluated at the background level.}  
\begin{equation}
    \text{d} s^2 = \bar{g}_{\mu\nu} \text{d} x^\mu \text{d} x^\nu 
                 = - A(r) \text{d}t ^2 + \frac{\text{d} r^2}{B(r)} 
                   + C(r) \qty(\text{d}\theta^2 + \sin^2\theta \text{d}\varphi^2)\,.
    \label{metricBG}
\end{equation}
Notice that we kept a general function $C(r)$ for the area of the sphere of constant $r$ and $t$, because various solutions in the literature are written in this form.  

To satisfy the background symmetries of our spacetime, we will consider that the scalar field only depends on the radial coordinate, $\bar{\phi} = \phi(r)$ and the vector field takes the non-generic form
\begin{equation}
    \bar{A}_\mu = (\bar{A}_0(r), \bar{A}_1(r),0 ,0) \, .
    \label{vectorField}
\end{equation}
This function could also be decomposed into transverse and longitudinal modes using Helmholtz's theorem \cite{DeFelice2016}. But because we do not consider any mass term à la Proca $A_\mu A^\mu$, the component $\bar{A}_1$ will play no game in our equations. This background is not the most generic spherically symmetric spacetime\footnote{We are thankful to Julio Oliva for mentioning to us this point.}. In fact, we should only impose that the energy-momentum tensor $T_{\mu\nu}$ is invariant under the lie derivative along the generators of the group SO(3), or we could consider a less restrictive situation where the electromagnetic part of the energy-momentum tensor is spherically symmetric. Therefore we could consider $\bar{A}=\bar{A}_0(r)dt+\bar{P}\cos{\theta}d\varphi$ where $\bar{P}$ is the magnetic charge. We will consider in this paper only electric charge. It is a genuine restriction. 

Using our background metric, the kinetic terms for the scalar and vector fields take the simple form, $\bar{X} = -B {\phi'}^2/2$ and $\bar{F}=B{\bar{A}_0}'^{2}/(2A)$, where a prime represents a derivative with respect to $r$. 

Replacing the metric (\ref{metricBG}), the scalar and vector fields in the action (\ref{action}), we obtain the equations of motion (EOM) by varying the action with respect to the functions $X=\{A,B,C,\phi,\bar{A}_0\}$. We obtain the equations of motion $\cE_X = 0$
\begin{align}
    \cE_A & \equiv \frac{1}{C}\qty[ f_1 \qty(1 + \frac{B {C'}^2}{4 C}) - 
			B C f_1'' - \frac{1}{2} B' \qty( C f_1)' - B \qty( C' f_1)'  ] + 
			f_2 - 2 \bar{F} f_{2,F}		 \\
    \cE_B & \equiv \frac{1}{C}\qty[ f_1 \qty( 1 -\frac{B {C'}^2}{4 C}) -
			\frac{B A'}{2A} \qty(C f_1)' - B C' f_1' ] + 
			f_2 - 2 \bar{F} f_{2,F} - 2 \bar{X} f_{2,X} 	\\
    \cE_C & \equiv 2 f_2 - f_1\qty[ \sqrt{\frac{B}{A}} \qty(\sqrt{\frac{B}{A}} A' )' + 
            \sqrt{\frac{B}{C}} \qty(\sqrt{\frac{B}{C}} C' )'  + \frac{B A' C'}{2 A C}] - 
            \frac{[(f_1')^2 ABC]'}{AC f_1'} \\
    \cE_\phi & \equiv \p_r\qty[ \sqrt{AB} C f_{2,X} \, \phi'(r) ] + \sqrt{\frac{A}{B}} C \qty( f_{2,\phi} +  
                \frac12 f_{1,\phi} \, R )   \\
    \cE_{\bar{A}_0} & \equiv  Q'(r)\quad \text{with}~~ Q(r)= \sqrt{\frac{B}{A}} C \bar{A}_0' \, f_{2,F}
    \label{eq:A0}
\end{align}
where $'$ indicates a derivative wrt $r$. From the last equation follows that $Q(r)$ is constant; this expresses the existence of a conserved charge arising from the $U(1)$ symmetry.

It is useful to introduce the following variables to simplify the notations
\begin{equation}
    \cJ(r) = \sqrt{AB} C f_{2,X} \, \phi'(r) \; ,   \qquad\qquad 
    \cS(r) = \sqrt{\frac{A}{B}} C \qty( f_{2,\phi} +  \frac12 f_{1,\phi} \bar{R} ) \,,
    \label{JJ}
\end{equation}
hence, the equation of motion for the scalar field is written as $\cE_\phi = \cJ' + \cS=0$. 

\section{Perturbation formalism}
\label{formalism}
In this section, we will briefly summarize the formalism for perturbations around a static spherically symmetric spacetime. We consider a background metric $\bar{g}_{\mu\nu}$ and a perturbed metric $g_{\mu\nu}=\bar{g}_{\mu\nu}+h_{\mu\nu}$. The background metric will be described by the line element (\ref{metricBG}).

Under a reparametrization of the angles $(\theta,\varphi)$ into $R_2(\theta,\varphi),R_3(\theta,\varphi)$, the 10 components of the perturbed metric $g_{\mu\nu}$ transform as scalar, vector or tensor. For example, the component $g_{00}$ transforms as a scalar while $g_{0i}$, where $i=(2,3)$, transforms as a vector. In fact, $g_{0i}{\rm d}t{\rm d}x^i\rightarrow g_{0i} \partial_j R_i {\rm d}t{\rm d}x^j$. In summary, we have three scalars $(g_{00},g_{01},g_{11})$, two vectors of dimension two $(g_{0i},g_{1i})$ and a rank two tensor $(g_{ij})$, where $(i,j)$ take the values $(2,3)$. Each vector field has an orthogonal decomposition
\begin{align}
    v_i=\partial_i \omega +w_i\,,\quad \text{with}~~\nabla_i w^i=0
\end{align}
and the rank 2 tensor can be decomposed as
\begin{align}
    T_{ij}=\lambda g_{ij}+\nabla_{ij}\psi+\nabla_{(i}S_{j)}+\tau_{ij}\quad \text{with}~~\nabla_i S^i=\nabla_i\tau^i_{~j}=\tau^i_{~i}=0  
\end{align}
from which, we can conclude easily that in four dimensions $\tau_{ij}=0$ because it has 3 different components $(\tau_{22},\tau_{23},\tau_{33})$ and 3 restrictions $(\nabla_i\tau^i_{~j}=0,\tau^i_{~i}=0)$. In $D=4$, perturbations around spherically symmetric spacetime can be described by scalar and vector pertubations, which are also known as odd or axial for vector perturbations \cite{Regge57} and even or polar for scalar perturbations \cite{Zerilli70}. As we have seen, in higher dimensions, we have an additional tensor mode \cite{Gibbons:2002pq}.

Considering the symmetries of our background spacetime, all scalars can be decomposed in the basis of (scalar) spherical harmonics as
\begin{align}
    \Phi(t,r,\theta,\varphi) = \sum_{\ell,m}\Phi_{\ell m}(t,r)Y_\ell^m(\theta,\varphi)
\end{align}
and all vectors can be decomposed into a basis of vector spherical harmonics\footnote{The first term comes easily from $\nabla_i\omega$ where $\omega$ is decomposed as any scalar, while the second term comes from the curl-free component. As we know $(\vec{\nabla}\times \vec{A})^i=\epsilon^{ijk}\nabla_j A_k$, which in 2D gives $\omega^i=\epsilon^{ij}\nabla_j\psi$ with $\psi$ a scalar that can be decomposed in the basis of scalar spherical harmonics.}
\begin{align}
    v_i(t,r,\theta,\varphi) = \sum_{\ell,m}\Bigl[\omega_{\ell m}(t,r)\nabla_i Y_\ell^m(\theta,\varphi) 
                              + \bar{\omega}_{\ell m}(t,r)E_i^{~j}\nabla_jY_\ell^m(\theta,\varphi)\Bigr] \, ,
\end{align}
where $E_{ij}=\sqrt{\det \gamma}~ \epsilon_{ij}$, $\gamma_{ij}$ being the two-dimensional metric on the sphere and $\epsilon_{ij}$ being the Levi-Civita symbol with $\epsilon_{23}=1$. The functions $(\nabla_i Y_\ell^m,E_i^{~j}\nabla_j Y_\ell^m)$ define the basis of our vector spherical harmonics.

In summary, we have 3 vector perturbations
\begin{align}
    h_{0i} &=\sum_{\ell m}h_{\ell m}^{(0)}(t,r)E_{ij}\partial^jY_\ell^{m} \, , \label{eq:pertV1}\\
    h_{1i} &=\sum_{\ell m}h_{\ell m}^{(1)}(t,r)E_{ij}\partial^jY_\ell^{m} \, , \label{eq:pertV2}\\
    h_{ij} &= \frac12 \sum_{\ell m}h_{\ell m}^{(2)}(t,r) \Bigl[E_i^{~k}\nabla_{kj}Y_\ell^{m}+E_j^{~k}\nabla_{ki}Y_\ell^{m}\Bigr] \, ,
    \label{eq:pertV3}
\end{align}
and 7 scalar perturbations
\begin{align}
    h_{00} &= A(r)\sum_{\ell m}H_{\ell m}^{(0)}(t,r)Y_\ell^{m}              \, ,  \\
    h_{01} &= \sum_{\ell m}H_{\ell m}^{(1)}(t,r)Y_\ell^{m}                  \, ,  \\
    h_{11} &= \frac{1}{B(r)}\sum_{\ell m}H_{\ell m}^{(2)}(t,r)Y_\ell^{m}    \, ,  \\
    h_{0i} &= \sum_{\ell m}\beta_{\ell m}(t,r)\partial_i Y_\ell^{m}         \, ,  \\
    h_{1i} &= \sum_{\ell m}\alpha_{\ell m}(t,r)\partial_i Y_\ell^{m}        \, ,  \\
    h_{ij} &= \sum_{\ell m}\Bigl[K_{\ell m}(t,r)g_{ij}Y_\ell^{m}+G_{\ell m}(t,r)\nabla_{ij}Y_\ell^{m}\Bigr] \, .
\end{align}
These perturbations are not independent. We can eliminate four of them using the coordinate transformation $x^\mu\rightarrow x^\mu+\xi^\mu$ where $\xi^\mu$ are infinitesimal. This transformation can also be decomposed into scalar and vector perturbations, we have for the scalar part
\begin{align}
    \xi_0 &=\sum_{\ell m}T_{\ell m}(t,r)Y_\ell^m(\theta,\varphi)                 \, ,  \\
    \xi_1 &=\sum_{\ell m}R_{\ell m}(t,r)Y_\ell^m(\theta,\varphi)                 \, ,  \\
    \xi_i &=\sum_{\ell m}\Theta_{\ell m}(t,r)\partial_i Y_\ell^m(\theta,\varphi) \, ,
\end{align}
and for the vector part
\begin{align}
    \xi_i=\sum_{\ell m}\Lambda_{\ell m}(t,r)E_i^j\partial_j Y_\ell^m(\theta,\varphi) \, .
\end{align}
Under this transformation, the perturbed metric transforms according to $h_{\mu\nu}\rightarrow h_{\mu\nu}-2\nabla_{(\mu}\xi_{\nu)}$ and therefore, we have
\begin{align}
  h_{\ell m}^{(0)} & \rightarrow h_{\ell m}^{(0)} - \dot\Lambda_{lm}                         \, ,\label{Xiv1}    \\
  h_{\ell m}^{(1)} & \rightarrow h_{\ell m}^{(1)} - \Lambda_{lm}'+\frac{C'}{C}\Lambda_{lm}   \, ,\label{Xiv2}    \\
  h_{\ell m}^{(2)} & \rightarrow h_{\ell m}^{(2)} - 2\Lambda_{lm}                            \, ,
\end{align}
where a dot indicates a derivative wrt $t$. 
We see that for vector perturbations and for $\ell\geq 2$, the only full gauge fixing condition is the Regge-Wheeler gauge defined by $h_{\ell m}^{(2)}=0$ while for scalar perturbations we have 
\begin{align}
    H_{\ell m}^{(0)} & \rightarrow H_{\ell m}^{(0)}-2\frac{\dot T_{lm}}{A}+\frac{BA'}{A} R_{lm}             \, ,  \\
    H_{\ell m}^{(1)} & \rightarrow H_{\ell m}^{(1)}-\dot R_{lm}-T_{lm}'+\frac{A'}{A}T_{lm}                  \, ,  \\
    H_{\ell m}^{(2)} & \rightarrow H_{\ell m}^{(2)}-2 B R_{lm}'-B' R_{lm}                                   \, ,  \\
    \beta_{\ell m}^{(0)}  & \rightarrow \beta_{\ell m}^{(0)}-\dot\Theta_{lm}-T_{lm}                         \, ,  \\
    \alpha_{\ell m}^{(0)} & \rightarrow \alpha_{\ell m}^{(0)}-\Theta_{lm}'-R_{lm}+\frac{C'}{C}\Theta_{lm}   \, ,  \\
    K_{\ell m}^{(0)} & \rightarrow K_{\ell m}^{(0)}-\frac{BC'}{C} R_{lm}                                    \, ,  \\
    G_{\ell m}^{(0)} & \rightarrow G_{\ell m}^{(0)}-2 \Theta_{lm}                                           \, .
\end{align}
From which we see that 3 gauges (in the gravity sector) are possible, defined by 
$(G_{\ell m}=K_{\ell m}=\beta_{\ell m}=0)$, 
$(G_{\ell m}=H^{(0)}_{\ell m}=\beta_{\ell m}=0)$ or 
$(G_{\ell m}=\alpha_{\ell m}=\beta_{\ell m}=0)$. 
We will use in our calculations, the first gauge.

These perturbations should be supplemented by the perturbation of the scalar field and the vector field. 
The scalar field $\phi$ has only a scalar perturbation\footnote{which transforms under a gauge transformation as $\delta\phi_{\ell m}\rightarrow \delta\phi_{\ell m} - B \, \bar\phi' R_{\ell m}$. We will later use this condition.}
\begin{align}
    \phi(t,r,\theta,\varphi) = \bar\phi(r)+\sum_{\ell m}\delta\phi_{\ell m}(t,r)Y_\ell^m(\theta,\varphi) \, ,
\end{align}
while the vector field perturbation ($A_\mu=\bar A_\mu+\delta A_\mu$) can be decomposed into a scalar and a vector part.
The vector perturbation is
\begin{align}
    \delta A_0 = \delta A_1=0 \,,\qquad 
    \delta A_i=\sum_{\ell m}A^{(v)}_{\ell m}(t,r) E_{ij}\nabla^j Y_\ell^m(\theta,\varphi) \, ,
    \label{eq:pertAv}
\end{align}
while the scalar perturbation can be written as
\begin{align}
    \delta A_0 &= \sum_{\ell m} A_{\ell m}^{(0)}(t,r)Y_\ell^m(\theta,\varphi)           \, , \\
    \delta A_1 &= \sum_{\ell m} A_{\ell m}^{(1)}(t,r)Y_\ell^m(\theta,\varphi)           \, , \\
    \delta A_i &= \sum_{\ell m} A_{\ell m}^{(2)}(t,r) \nabla_i Y_\ell^m(\theta,\varphi) \, .
\end{align}
Because of the gauge freedom associated to the vector field\footnote{We define the transformation $A_0\rightarrow A_0-\sum_{\ell m}\dot A_{\ell m}^{(2)}Y_\ell^m$, $A_1\rightarrow A_1-\sum_{\ell m} A_{\ell m}^{'(2)}Y_\ell^m$ and $A_i\rightarrow A_i-\sum_{\ell m} A_{\ell m}^{(2)} \nabla_i Y_\ell^m$. We have for example $A_{\ell m}^{(0)}\rightarrow A_{\ell m}^{(0)}-\dot A_{\ell m}^{(2)}$, that we will redefine as $A_{\ell m}^{(0)}$ without any loss of generality. It is important that scalar perturbations remain scalar after this redefinition.}, we can set $A_{\ell m}^{(2)}$ to zero.

We will see later that for $\ell=(0,1)$ some of the perturbations are identically zero which allows us to use other gauges and simplify the problem.
The formalism summarized in this section applies to scalar and vector perturbations when $\ell\geq 2$. 

In summary, we have two type of modes which can be studied separately at the linear order unless we have parity-violating terms \cite{Motohashi11} such as Pontryagin density $*R R$. 
Also for any spherically symmetric background, the final equations of perturbations are independent of the order $m$ \cite{Chandrasekhar98}. 
Therefore, without any loss of generality, we will assume $m=0$.

\section{Odd-parity perturbations}
Over the static and spherically symmetric background (\ref{metricBG}), we will consider small perturbations of the fields defined by eqs.(\ref{eq:pertV1}, \ref{eq:pertV2}, \ref{eq:pertV3}, \ref{eq:pertAv})

Since modes with different $\ell$ evolve independently, we focus on a specific mode and omit the index. 

\subsection{Second-order action for higher multipoles, \texorpdfstring{$\ell \geq 2$}{l>=2}}
In this section we will focus in higher multipoles ($\ell \geq 2$), since dipole mode $\ell = 1$ require a special treatment. 
Expanding the action (\ref{action}) to second order in perturbations\footnote{The first order perturbations vanish on-shell.} and performing integration over the sphere $(\theta,\varphi)$, we find
\begin{equation}
    S^{(2)}_{\mbox{\tiny odd}} = \frac{2\ell + 1}{4\pi}\int {\rm d} t\,{\rm d} r\,\cL^{(2)}_{\mbox{\tiny odd}} \,,
    \label{action-init}
\end{equation}
where\footnote{We have renamed for simplicity of notations $h^{(0)}_{\ell m}\rightarrow h_0$, $h^{(1)}_{\ell m}\rightarrow h_1$, $A^{(v)}_{\ell m}\rightarrow A_v$}
\begin{equation}
    \begin{aligned}
    \cL^{(2)}_{\mbox{\tiny odd}} & = 
        a_1 h_0^2 + a_2 h_1^2 
        + a_3 \qty[\dot{h}_1^2 + h_0'^2 + 2 \frac{C'}{C} h_0 \dot{h}_1 - 2 h_0'  \dot{h}_1 + 2 a_4\qty(h_0' - \dot{h}_1)A_v] 
        \\ & \qquad 
        + a_5 h_0 A_v + a_6 A_v^2 + a_7 \dot{A}_v^2 + a_8 A_v'^2
    \end{aligned}
    \label{lagrangian}
\end{equation}
where the coefficients are given by 
\begin{equation}
    \begin{array}{ll}
    a_1 = \displaystyle\frac{\l}{4 C} \Bigl[\frac{{\rm d}}{{\rm d} r}\left( C'\sqrt{\frac{B}{A}}f_1\right) +\frac{(\l-2)f_1}{\sqrt{AB}} + \frac{2C}{\sqrt{AB}}  {\cal E}_A \Bigr] \, , \qquad &
    a_5 = \displaystyle\frac{\l}{C}\qty( {\cal E}_{\bar{A}_0} - \frac{C'}{C} Q)	                            \\
    a_2 = \displaystyle- \frac{\l}{2} \sqrt{AB} \left[ \frac{(\l-2)f_1}{2 C} + {\cal E}_B \right]   \,, 	& 
    a_6 = \displaystyle-\frac{\l^2 \bar{A}_0'}{4C^2} \frac{Q}{\bar{F}}                                      \\
    a_3 = \displaystyle\frac{\l}{4} \sqrt{\frac{B}{A}} f_1  \,,                                             & 
    a_7 = \displaystyle\frac{\l}{2 BC \bar{A}_0' } Q                                                        \\
    a_4 = \displaystyle2\sqrt{\frac{A}{B}} \frac{Q}{C f_1} \,,                                              & 
    a_8 = \displaystyle-\frac{\l A}{2 C \bar{A}_0'} Q
    \end{array}
    \label{odd-coefficients}
\end{equation}
and $\l = \ell(\ell +1)$. On-shell, ${\cal E}_A={\cal E}_B = {\cal E}_{\bar{A}_0} =0$ which allows us to simplify these coefficients. 
Notice that for $\ell=1$, we have $a_2=0$ on-shell, this is why this mode has to be studied separately.
The term inside square brackets in (\ref{lagrangian}) can be rewritten in a more convenient way such that the Lagrangian is simplified to
\begin{equation}
    \cL^{(2)}_{\mbox{\tiny odd}} = b_1 h_0^2 + a_2 h_1^2 + a_3 \qty[ \dot{h}_1 - h_0' + \frac{C'}{C} h_0 - a_4 A_v ]^2 
                                   + \qty( a_6 - a_3 a_4^2) A_v^2  + a_7 \dot{A}_v^2 + a_8 A_v'^2 + b_2 h_0 A_v
    \label{Lag_L2}
\end{equation}
where we have defined, $b_1 = a_1 - (C'a_3)'/C$ and $b_2 = a_5 + 2 a_3 a_4 C'/C\equiv \lambda {\cal E}_{\bar{A}_0}/C$. 
On-shell, the last coefficient vanishes, $b_2 = 0$.

Since no time derivative of $h_0$ appears in the Lagrangian, variation with respect to it, yields a constraint equation. However, because of the presence of $h'_0$ in the action, this constraint results in a second-order ordinary differential equation which cannot be immediately solved for $h_0$. To overcome this obstacle, we follow the procedure described in \cite{DeFelice11, Ganguly18}, we introduce an auxiliary field $q(t,r)$ and define the following Lagrangian
\begin{equation}
    \cL^{(2)}_{\mbox{\tiny odd}} = b_1 h_0^2 + a_2 h_1^2 + a_3 \qty[2q\qty(\dot{h}_1 - h_0' + \frac{C'}{C} h_0 - a_4 A_v) - q^2]
                                   + \qty( a_6 - a_3 a_4^2) A_v^2  + a_7 \dot{A}_v^2 + a_8 A_v'^2 \, . 
    \label{Lag-q}
\end{equation}
We can easily check that by substituting the EOM for $q(t,r)$ into Eq. (\ref{Lag-q}), we recover the original Lagrangian. 
Now, varying (\ref{Lag-q}) with respect to $h_0$ and $h_1$ leads to
\begin{align}
    b_1 h_0 + \qty[ (a_3 q )' + a_3 q  \frac{C'}{C}] = 0    &   \qquad \longrightarrow \qquad
        h_0 = - \frac{(a_3 q C)'}{C b_1}    \\
    a_2 h_1 - a_3 \dot{q} = 0 & \qquad \longrightarrow \qquad h_1 = \frac{a_3}{a_2}\dot{q}
\end{align}
these expressions relate the metric elements $h_0$ and $h_1$ to the auxiliary field $q$. 
Once $q(t,r)$ is known, $h_0$ and $h_1$ are easily obtained, and thus all vector perturbations in the Regge-Wheeler gauge are obtained.
Substituting these expressions into the Lagrangian and performing integration by parts one finds
\begin{equation}
    \cL^{(2)}_{\mbox{\tiny odd}} =  
        \alpha_1 \dot{q}^2 + \beta_1 q'^2 + \gamma_1 q^2 + 
        \alpha_2 \dot{A}_v^2 + \beta_2 A_v'^2 + \gamma_2 A_v^2 + 
        \sigma A_v q  \, ,
        \label{eq:prevector}
\end{equation}
which explicitly shows that taking into account the vector sector, we only have two degrees of freedom, one associated with gravitational perturbation and another related to vector perturbation of the electromagnetic field. 
The coefficients of the above Lagrangian, in terms of the variables (\ref{odd-coefficients}), are 
\begin{equation}
    \begin{aligned}
    \alpha_1 & = -\frac{a_3^2}{a_2}    \, , \hspace{2.2cm} 
    \alpha_2   = a_7                   \, , \hspace{2.2cm} 
    \sigma     = - 2 a_3 a_4           \, , \\
    \beta_1 &  = -\frac{a_3^2}{b_1}    \, , \hspace{2.2cm} 
    \beta_2    = a_8	               \, , \hspace{2.2cm} 
    \gamma_2   = a_6 - a_3 a_4^2       \, , \\
    \gamma_1 & = \frac{a_3}{b_1^2 C^2} \left[-C b_1' (a_3 C)' + b_1 \left[C (a_3 C)'' -2 C' (a_3 C)' \right]-b_1^2 C^2\right] \, .
\end{aligned}
\end{equation}
We see that in the absence of electromagnetic perturbations, the Lagrangian (\ref{eq:prevector}) becomes $\cL^{(2)}_{\mbox{\tiny odd}} =\alpha_1 \dot{q}^2 + \beta_1 q'^2 + \gamma_1 q^2$, from which we should at least impose the no-ghost condition $\alpha_1>0$ translating into $a_2<0$ and therefore $f_1>0$. 
Similarly, in the absence of the gravitational perturbations, we have $\cL^{(2)}_{\mbox{\tiny odd}} = \alpha_2 \dot{A}_v^2 + \beta_2 A_v'^2 + \gamma_2 A_v^2$ from which we impose the condition $\alpha_2>0$ which means $Q/\bar{A}_0'>0$ and using eq.(\ref{eq:A0}) gives $f_{2,F}>0$.

\subsection{Master equation}
To arrive at our final result, it is convenient to rescale the variables as
\begin{equation}
    q(t,r)   = \sqrt{\frac{A}{f_1 BC}} V_g (t,r)        \, , \qquad 
    A_v(t,r) = \frac{V_e(t,r)}{\sqrt{2(\l -2)f_{2,F}}}  \, ,
    \label{Vector-Change}
\end{equation}
where $(V_g,V_e)$ represent respectively the vector perturbations associated to the gravitational and electromagnetic sector. 
Notice that we have assumed the previous conditions $f_1 > 0$ and  $f_{2,F} > 0$. 
Finally, introducing the tortoise coordinate, $\dd r = \sqrt{AB} \, \dd r_*$ we find
\begin{equation}
    S^{(2)}_{\mbox{\tiny odd}} = \frac{\ell(\ell +1)(2\ell + 1)}{16 \pi (\ell + 2)(\ell -1)}
        \int \dd t \dd r_* \qty[
        \qty(\frac{\partial \Psi_i}{\partial t})^2 - \qty(\frac{\partial \Psi_i}{\partial r_*})^2 - V_{ij} \Psi_i \Psi_j ]\, ,
    \label{S2-action}
\end{equation}
where $\vec{\Psi} = (V_g,\, V_e)^t$, and $V_{ij}$ are the components of a $2 \times 2$ symmetric matrix and are given, after substituting relations (\ref{odd-coefficients}), by
\begin{align}
    V_{11} & = \qty(\l - 2)\frac{A}{C} - \p_{r_*} S_1 + S_1^2       \, , ~~\quad\qquad~ 
    S_1(r) =\frac{\sqrt{AB}}{2}\qty(\frac{C'}{C}+\frac{f_1'}{f_1})  \, ,                               
    \label{Va_potential} \\
    V_{22} & = \frac{\lambda  A}{C} + G(r) - \p_{r_*} S_2 + S_2^2   \, ,\qquad\quad\;
    S_2(r) = -\frac{\sqrt{AB}}{2}\frac{f_{2,F}'}{f_{2,F}}           \, ,
    \label{Vb_potential} \\
    V_{12} & = V_{21} = \sqrt{(\l -2)\frac{A}{C} G(r)}              \, , ~ \qquad \qquad 
    G(r) = 2\frac{A}{C^2}\frac{Q^2}{f_1 f_{2,F}}                    \, .
    \label{Vc_potential}
\end{align}
The introduction of the functions $(S_1,S_2)$ will be more transparent in the stability analysis section. 
Variation of the action (\ref{S2-action}) with respect to $\Psi$ gives us a wave-like equation
\begin{equation}
    -\frac{\partial^2 \vec{\Psi}}{\partial t^2} + \frac{\partial^2 \vec{\Psi}}{\partial r_*^2} - \mathbf{V}\vec{\Psi} = 0 \, .
    \label{System-Eq}
\end{equation}
Both modes propagate at the speed of light.
Since matrix $\mathbf{V}$ is not diagonal, the previous equation is a set of two coupled differential equations. But the system can be diagonalized because the eigenvalues are independent of the radial coordinate\footnote{We could verify it for various theories but not generically.}.

\subsection{Stability analysis}
\label{stability1}
Stability means that no perturbation grows unbounded in time. 
For that, we will Fourier transform our variables $(\vec{\Psi}\rightarrow e^{-i\omega t} \vec{\Psi})$ such as eq.(\ref{System-Eq}) is recast as
\begin{align}
    \mathcal{H}\vec{\Psi}=\omega^2\vec{\Psi}
\end{align}
where $\mathcal{H}=-\partial_{r_*}^2+\mathbf{V}$. 
The frequency $\omega^2$ appears as the eigenvalues of the operator $\mathcal{H}$. 
Unstable modes are equivalent to  purely imaginary modes $\omega^2<0$ (see e.g. \cite{Ganguly18}), therefore the stability of the spacetime is related to the positivity of the operator $\mathcal{H}$, namely that $\mathcal{H}$ has no negative spectra. 
To prove the stability, let us define the inner product 
\begin{align}
    (\vec{\psi},\vec{\xi})=\int {\rm d}r_* \Bigl[\bar{\psi}_1 \xi_1 + \bar{\psi}_2 \xi_2 \Bigr] \, ,
\end{align}
where $\vec{\psi}=(\psi_1,\psi_2)^T$ and $\vec{\xi}=(\xi_1,\xi_2)^T$. 
Stability means that the operator $\mathcal H$ is a positive self-adjoint operator in $L^2(r_*)$, the Hilbert space of square integrable functions of $r_*$. 
Therefore, we need to prove the positivity defined as
\begin{align}
    \forall \chi\,,\quad (\vec{\chi},\mathcal{H}\vec{\chi})>0 \, .
\end{align}
This condition will imply that given a well-behaved initial data, of compact support, $\chi$ remains bounded for all time. 
This is a sufficient condition. 
The rigorous and complete proof of the stability related to equations of the form $(\ref{System-Eq})$ can be found in \cite{wald1,wald2} using spectral theory.

We have
\begin{align}
    (\vec{\chi},\mathcal{H}\vec{\chi}) &=\int {\rm d}r_*\Bigl[\bar{\chi}_1\Bigl(-\partial_{r_*}^2\chi_1+V_{11}\chi_1+V_{12}\chi_2\Bigr)+\bar{\chi}_2\Bigl(-\partial_{r_*}^2\chi_2+V_{22}\chi_2+V_{12}\chi_1\Bigr)\Bigr]\nonumber\\
    &=\int {\rm d}r_*\Bigl[\Big|\frac{{\rm d}\chi_1}{{\rm d}r_*}\Bigr|^2+\Big|\frac{{\rm d}\chi_2}{{\rm d}r_*}\Bigr|^2+V_{12}\Bigl(\bar{\chi}_1\chi_2+\chi_1\bar{\chi}_2\Bigr)+V_{11}|\chi_1|^2+V_{22}|\chi_2|^2\Bigr]\nonumber\\
    & = \int {\rm d}r_*\Bigl[\Big|\frac{{\rm d}\chi_1}{{\rm d}r_*}+S_1\chi_1\Bigr|^2+\Big|\frac{{\rm d}\chi_2}{{\rm d}r_*}+S_2\chi_2\Bigr|^2+(\lambda-2)\frac{A}{C}|\chi_1|^2+(\lambda \frac{A}{C}+G)|\chi_2|^2\nonumber\\
    &\quad +\sqrt{(\lambda-2)\frac{A}{C}G}\Bigl(\bar{\chi}_1\chi_2+\chi_1\bar{\chi}_2\Bigr)\Bigr]\nonumber\\
    & = \int {\rm d}r_*\Bigl[\Big|\frac{{\rm d}\chi_1}{{\rm d}r_*}+S_1\chi_1\Bigr|^2+\Big|\frac{{\rm d}\chi_2}{{\rm d}r_*}+S_2\chi_2\Bigr|^2+\Bigl|\sqrt{(\lambda-2)\frac{A}{C}}\chi_1+\sqrt{G}\chi_2\Bigr|^2+\lambda \frac{A}{C}|\chi_2|^2\Bigr]\nonumber\\
    &\geq 0  \, .
\end{align}
where $\bar{\chi}_i$ ($i=1,2$) represent the complex conjugate of $\chi_i$ and should not be confused with background quantities.
In the second line, we have neglected the boundary term $\bar{\chi}_1\partial_{r_*}\chi_1+\bar{\chi}_2\partial_{r_*}\chi_2$ coming from the integration by parts, because we assumed $\chi_1$ and $\chi_2$ to be smooth functions of compact support, while in the third line we have neglected the boundary term $S_1|\chi_1|^2+S_2|\chi_2|^2$. 

Therefore, we conclude that the black hole is stable under vector perturbations for $\ell\geq 2$ if the no-ghost condition is satisfied, namely\footnote{It was also noticed that in Horndeski theory, the black hole is stable if the no-ghost and hyperbolicity conditions are satisfied \cite{Ganguly18}. Notice that for Lovelock black holes, the no-ghost and hyperbolicity conditions could be satisfied while the black hole is unstable \cite{Takahashi:2010gz}.}
\begin{align}
    f_1(\phi)>0\,, \quad    f_{2,F}(\phi,X,F) >0.
    \label{VecStability}
\end{align}
Notice that these conditions are not equivalent to energy conditions (see Appendix \ref{energy}).

\subsection{Stability of dipole perturbation, \texorpdfstring{$\ell = 1$}{l=1}}
We should first mention that for $\ell =0$, spherical harmonics are constant and therefore the three vector perturbations of the metric and the vector perturbation of the electromagnetic field are identically zero as we can see from their definition (\ref{eq:pertV1}, \ref{eq:pertV2}, \ref{eq:pertV3}, \ref{eq:pertAv}).

For the dipole perturbation, $\ell =1$ and assuming $m=0$ because the final result is independent of the azimuthal angle, we have $Y_1^0\propto \cos{\theta}$, therefore we have $h_{ij}=0$ where $(i,j)=\{1,2\}$. 
We don't need to use the Regge-Wheeler gauge to eliminate this perturbation. 
We have seen that
\begin{align}
    h_0 &\rightarrow h_0-\dot\Lambda                    \, , \\
    h_1 &\rightarrow h_1-\Lambda'+\frac{C'}{C}\Lambda   \, .
\end{align}
Therefore, we can use this gauge freedom to fix $h_1=0$ but we have a residual gauge freedom defined as $\Lambda\rightarrow\Lambda +f(t)C(r)$.

Let us consider the Lagrangian (\ref{Lag_L2}) which is valid also for $\ell = 1$. 
After imposing the background equations, we get $a_2 = b_1 = 0$; and therefore\footnote{We will fix $h_1=0$ after using the constraint derived from its variation because the gauge is not totally fixed by this condition.}
\begin{equation}
    \cL^{(2)}_{\tiny \mbox{odd}, \, \ell=1} = a_3 \qty[ \dot{h}_1 - h_0' + \frac{C'}{C} h_0 - a_4 A_v ]^2 
        + \qty( a_6 - a_3 a_4^2) A_v^2  + a_7 \dot{A}_v^2 + a_8 A_v'^2  \, .
    \label{eq:L21}
\end{equation}
Variation of this action with respect to $h_1$ and $h_0$ gives us
\begin{equation}
    \dot{\cE} = 0 \, , \qquad\qquad 
    (C \cE)'  = 0 \, ,
    \label{E-eqatuions}
\end{equation}
where we have defined
\begin{equation}
    \cE = a_3 \qty(\dot{h}_1 - h_0' + \frac{C'}{C} h_0 - a_4 A_v)  \,,
\end{equation}
which solution is given by $\cE = \mathcal{J}/C(r)$, where $\mathcal{J}$ is an integration constant\footnote{We hope that this will not confuse the reader because of the function defined in eq.(\ref{JJ}).}. 
Considering our gauge freedom, we fix $h_1=0$, and integrating the previous equation, we obtain 
\begin{align}
    h_0(t,r)=-\mathcal{J} C(r)\int\frac{{\rm d}r}{C(r)^2a_3}-C(r)\int \frac{a_4}{C(r)}A_v {\rm d}r+F(t)C(r)  \, ,
\end{align}
where $F(t)$ is a constant of integration. 
This last term can be eliminated with the help of the residual gauge freedom, giving finally
\begin{align}
    h_0(t,r)=-2\mathcal{J} C(r)\int\frac{1}{C^2(r)f_1}\sqrt{\frac{A}{B}}{\rm d}r 
        - 2Q C(r)\int \frac{A_v}{C(r)^2 f_1}\sqrt{\frac{A}{B}} {\rm d}r  \, .
\end{align}
The variation of (\ref{eq:L21}) wrt $A_v$ gives
\begin{align}
    a_7 \ddot A_v+(a_8 A_v')'-(a_6-a_3a_4^2)A_v+a_4\frac{\mathcal{J}}{C(r)}=0  \, .
\end{align}
Defining a new variable $a_v=\sqrt{f_{2,F}} A_v$ and using the tortoise coordinate, we find
\begin{align}
    -\frac{\partial^2 a_v}{\partial t^2}+\frac{\partial^2 a_v}{\partial r_*^2}-V_{22} a_v-\frac{2Q\mathcal{J}A}{C^2f_1\sqrt{f_{2,F}}}=0
    \label{eq:finalL21}
\end{align}
where $V_{22}$ is the coefficient defined in eq.(\ref{Vb_potential}) for $\ell=1$, i.e. $\lambda=2$. 

The solution of eq.(\ref{eq:finalL21}) is the sum of a particular solution that we can consider as a function of $r$ only and a homogeneous solution. 
In order to understand this solution, let us consider the simple case where we have perturbed Reissner-Nordström in general relativity. 
For that we take $f_1=2$, $f_2=4F$ and $Q=4q$, from which we get $A=B=1-2M/r+q^2/r^2$ and $C=r^2$. 
For this spacetime, a particular solution of eq.(\ref{eq:finalL21}) is $a_v=-Jq/6Mr$ which gives $A_v=-Jq/12Mr$ and therefore $h_0=-\frac{\mathcal{J}}{6M}\Bigl(\frac{-2M}{r}+\frac{q^2}{r^2}\Bigr)$ which means 
\begin{align}
    h_{03}=-\frac{\mathcal{J}}{6M}\Bigl(\frac{-2M}{r}+\frac{q^2}{r^2}\Bigr)\sin^2{\theta}  \, , 
\end{align}
in other words, the Kerr-Newman solution at the first order in $J$ if we define $\mathcal{J}=-12 J$. 
Therefore, the particular solution of eq.(\ref{eq:finalL21}) should describe a slow rotating black hole while the homogeneous solution describes the propagation of the electromagnetic field in our spacetime \cite{Moreno:2002gg}. 
This propagation is stable because of the potential $V_{22}$ which can be easily deformed to a positive potential using the function $S_2(r)$. 
In fact, as we have shown in Sec.(\ref{stability1}), the operator associated is definite positive. 

In conclusion, we have shown that for any theory described by the action (\ref{action}) and considering the conditions $f_1>0$ and $f_{2,F}>0$, the static spherically symmetric spacetime described by a generic metric (\ref{metricBG}) is stable under vector perturbations.

\section{Even-parity perturbations}

In this section, we will study the scalar (even, polar) perturbations of the action (\ref{action}). We follow the formalism that we have summarized in Sec.(\ref{formalism})

\subsection{Second-order action for higher multipoles, \texorpdfstring{$\ell \geq 2$}{l>=2}}
Following exactly the same procedure, we arrive at 
\begin{equation}
S^{(2)}_{\mbox{\tiny even}} = \frac{2\ell + 1}{2\pi}\int {\rm d} t\,{\rm d} r\,\cL^{(2)}_{\mbox{\tiny even}}\,,
\end{equation}
where the second-order Lagrangian is
\begin{equation}
    \begin{aligned}
    \cL^{(2)}_{\mbox{\tiny even}} & = 
    a_1 H_0^2 + H_0 \left[
    a_2 H_2' + \l a_3 \alpha' + (a_4 + \l a_5) H_2 + \l a_6 \alpha + (a_7 + \l a_8) \d\phi + a_9 \d\phi '  \right. \\
    & \left. 
    + a_{10} \d\phi''  + a_{11} (A_0' - \dot{A}_1) \right] +
    (b_1 + \l b_2) H_1^2 + H_1 \qty[
    b_3 \dot{H}_2 + \l b_4 \dot{\alpha} + b_5 \dot{\d\phi} + b_6 \dot{\d\phi}' ] \\
    & 
    + c_1 H_2^2 + \l c_2 \alpha^2 + \l c_3 H_2 \alpha + \l c_4 \dot{\alpha}^2 +
    (c_5 + \l c_6) H_2 \d\phi + c_7 H_2 \d\phi ' + c_8 \dot{H}_2 \dot{\d\phi} \\
    & 
    + \l c_9 \alpha \d\phi + \l c_{10} \alpha \d\phi' + c_{11} H_2 (A_0' - \dot{A}_1) + \l c_{12} \alpha A_0 
    + e_1 \dot{\d\phi}^2 + e_2 \d\phi '^2  \\ 
    & 
    + (e_3 + \l e_4) \d\phi^2 + e_5 \d\phi (A_0' - \dot{A}_1) + e_6 \d\phi ' (A_0' - \dot{A}_1) \\
    &
    + d_1 (A_0' - \dot{A}_1)^2 + \l d_2 A_0^2 + \l d_3 A_1^2 \, ,
\end{aligned}
\label{L-evenperturbation}
\end{equation}
here $a_i$, $b_i$, $c_i$, $d_i$ and $e_i$ are all functions of $r$ only and their expressions are given in the Appendix \ref{EvenL-coe}. 
In what follows, we will reduce this Lagrangian and rewrite it in terms of three variables, representing the remaining three DOF of the theory. 
The variation of the action with respect to $A_0$ and $A_1$ gives
\begin{align}
    \l c_{12} \alpha + 2\l d_2 A_0 -\p_r \qty[a_{11} {H_0}+c_{11} {H_2} + {e_5} \delta \phi + {e_6} \delta \phi' + 2 {d_1} ({A_0}' - \dot{A}_1)] &= 0  \, ,\\
    2 \l  d_3 A_1 + \p_t \qty[ a_{11} {H_0}+c_{11} {H_2} + {e_5} \delta \phi + {e_6} \delta \phi' + 2 {d_1} ({A_0}' - \dot{A}_1) ]               &= 0  \, .
\end{align}
Introducing the new variable\footnote{which will be associated to the scalar part perturbation of the electromagnetic field.} defined as
\begin{equation}
    S_e(t,r) = a_{11} {H_0}+c_{11} {H_2} + {e_5} \delta \phi + {e_6} \delta \phi' + 2 {d_1} ({A_0}' - \dot{A}_1) \, ,
    \label{eq:H-def}
\end{equation}
we obtain
\begin{equation}
    A_0 = \frac{1}{2 \l {d_2}} \qty(S_e' - \l c_{12} {\alpha})      \, , \qquad\qquad
    A_1 = -\frac{\dot{S_e}}{2 \lambda {d_3}}\, .
    \label{h4-h5-def}
\end{equation}
Let us notice from Lagrangian (\ref{L-evenperturbation}) that $H_1$ has no derivatives, so we can integrate out this non-propagating field by using its own equation of motion
\begin{equation}
    H_1 = -\frac{1}{2 \lambda {b_2}}\qty[{b_3} \dot{H}_2 + \l {b_4} \dot{\alpha} + {b_5} \dot{\delta \phi} + {b_6} \dot{\delta \phi}']\, .
\end{equation}
Notice that the coefficient $b_1$ is zero on-shell, so we have dropped this term in the previous equation.
At this point, the Lagrangian (\ref{L-evenperturbation}) only depends on the variables $H_0$, $H_2$, $\alpha$, $S_e$ and $\d\phi$. 
Variation of the action with respect to $H_0$ gives us
\begin{equation}
    \begin{aligned}
    \cE_1 \equiv & \; 2 a_1 H_0 +
    a_2 H_2' + \l a_3 \alpha' + (a_4 + \l a_5) H_2 + \l a_6 \alpha + \\ & 
    (a_7 + \l a_8) \d\phi + a_9 \d\phi ' + a_{10} \d\phi''  + a_{11} (A_0' - \dot{A}_1) = 0 \, .
    \end{aligned}
\end{equation}
Taking the combination $\cE_1 - a_{11} / (2 d_1) S_e(t,r)$, and using the identity $2a_1 - a_{11}^2/2d_1 = 0$, we obtain a relation that sets a constraint for the other fields, but this is not an algebraic constraint; however, in order to resolve this issue, we perform a field redefinition and use
a new variable\footnote{which will be associated to the scalar part perturbation of the gravitational field.} $S_g$ defined by
\begin{equation}
H_2 = \frac{1}{a_2}\qty[S_g(t,r) - a_{10} \delta \phi' - \l a_3 {\alpha}] \, .
\end{equation}

Using this relation, the constraint becomes an algebraic equation for $\alpha$, which can be solved 
\begin{equation}
\alpha = \frac{a_{3}}{\l {b_4}}T(r) \left[S_g' + \l \frac{a_{5}}{a_{2}}S_g
	  + \left(P(r) - \l\frac{ a_{10} a_{5}}{a_{2}}\right) \delta \phi' 
	  +\frac{a_{11} S_e}{2 {d_1}} 
	  + (P'(r) + \l a_{8}) \delta \phi 
	  \right]
\end{equation}
where $T$ and $P$ are given by
\begin{equation}
    \begin{aligned}
    T(r) & = \frac{2}{A f_1 \sqrt{AB}}\qty(\frac{C'}{C} - \frac{A'}{A} - \frac{2\l f_1}{B(Cf_1)'})^{-1}  \, ,
    \\
    P(r) & = \sqrt{B} C \left[A \qty(\frac{f_{1,\phi}}{\sqrt{A}})' + \sqrt{A} \phi' {f_{2,X}}\right] \, .
    \end{aligned}
    \label{eq:TP-def}
\end{equation}

Replacing this definition into $\cE_1$, we can rewrite $H_0$ in terms of the other variables and their derivatives (this is a large expression, and at this point, it is unnecessary to write it). 
Substituting this relations into (\ref{L-evenperturbation}) we obtain a Lagrangian that only depends on variables $S_g(t,r)$, $S_e(t,r)$ and $\d\phi(t,r)$ given by
\begin{equation}
    \begin{aligned}
    \cL_{\mbox{\tiny even}}^{(2)} &= 
    	{\alpha_1} \dot{S_g}^2 + {\beta_1}{S_g'}^2 + {\gamma_1} S_g^2 + 
    	{\alpha_2} \dot{S_e}^2 + {\beta_2}{S_e'}^2 + {\gamma_2} S_e^2 +
    	{\alpha_3} \dot{\d\phi}^2 + {\beta_3}{\d\phi'}^2 + {\gamma_3} \d\phi^2 +                                   \\ & \quad
    	{\sigma_1} S_g S_e + {\sigma_2} {S_g'}S_e  + {\sigma_3} {S_g'}{S_e'} + {\sigma_4} \dot{S_g}\dot{S_e} + 
    	{\eta_1} S_g \d\phi + {\eta_2} {S_g'}\d\phi  + {\eta_3} {S_g'}{\d\phi'} + {\eta_4} \dot{S_g}\dot{\d\phi} + \\ & \quad
    	{\nu_1} S_e{\d\phi} + {\nu_2} S_e{\d\phi}' + {\nu_3} S_e' {\d\phi}' + \nu_4 \dot{S_e}\dot{\d\phi}           \, .
    \end{aligned}
    \label{EvenL2-Lag}
\end{equation}
The above Lagrangian can be rewritten in matrix form
\begin{equation}
    \cL_{even}^{(2)} = 
    K_{ij} \dot{\chi}_i \dot{\chi}_j - L_{ij} \chi'_i \chi'_j  + D_{ij} \chi'_i \chi_j  + M_{ij} \chi_i \chi_j \, ,
    \label{eq:Even-Lag-MatrixForm}
\end{equation}
where $\chi = (S_g, \;S_e,\;\d\phi)^T$ and $(K_{ij},L_{ij},M_{ij})$ are symmetric matrices, while $D_{ij}$ is antisymmetric\footnote{All these matrices are given in a Mathematica file \cite{mathematica}}. 
The no-ghost condition requires the matrix $K$ to be positive definite; for this purpose, we employ the Sylvester's criterion, which gives
\begin{equation}
    K_{11} > 0                  \, ,\qquad\quad 
    K_{11}K_{22} - K_{12}^2 > 0 \, , \qquad\quad 
    \det(K_{ij}) > 0 \,.
    \label{eq:Sylvester-criterion}
\end{equation}
We obtain that the first two conditions reduce to
\begin{align}
    & K_{11} = \frac{ AT^2 }{\lambda C} \sqrt{\frac{A}{B}} \left(\lambda P_1 - f_1 + \frac{BC}{A}\bar{A}_0'^2 f_{2,F}\right) > 0  \, , \label{K11-Cond}\\ 
    & K_{11}K_{22} - K_{12}^2 = \frac{A}{B C}\qty(\frac{T}{2\l})^2 \frac{\lambda P_1 - f_1}{f_{2,F}}  > 0 \label{Inner-DetCond} \, .
\end{align}
where we defined
\begin{align}
    P_1(r) = \frac{B \Xi}{2 A C f_1^2}  \qty[\frac{A C^2 f_1^4}{\Xi ^2 B}]' \,, \qquad 
    \text{and} \qquad 
    \Xi=(Cf_1)' \, .
\end{align}

Imposing the conditions that we obtained in the odd-parity sector, namely $f_1 > 0$ and  $f_{2,F} > 0$, we get that the last term in (\ref{K11-Cond}) is always positive, so the previous two relations are reduced to a single constraint: $\lambda P_1 - f_1 > 0$. 
Finally, the third condition is given by
\begin{equation}
    \det(K_{ij}) = \sqrt{\frac{A}{B}}\qty(\frac{ T C'}{4 \l C \phi'})^2 \frac{(\l -2)f_1}{B \,  f_{2,F}} (2 P_1 - f_1) > 0 \, ,
\end{equation}
which is satisfied if and only if 
\begin{equation}
    2 P_1 - f_1 > 0 \, .
    \label{Stability-Condition}
\end{equation}
Clearly, if Eq. (\ref{Stability-Condition}) is satisfied then conditions (\ref{K11-Cond}) and (\ref{Inner-DetCond}) are satisfied automatically, given that $\ell \geq 2$.
We recover the no-ghost condition computed in \cite{Kobayashi2014} for a general scalar-tensor theory. 

Using the background equations, we find
\begin{align}
    2 P_1 - f_1  = \frac{C^2f_1 {\phi}'^2}{\Xi^2}\Bigl(3 {f_1}^2_{,\phi}+2 f_1{f_2}_{,X}\Bigr)
\end{align}
and assuming stability of odd-parity perturbations, viz. $f_1>0$, we obtain
\begin{align}
   3 {f_1}^2_{,\phi}+2 f_1{f_2}_{,X} >0 \, .
\end{align}

\subsection{Speed of propagation of scalar perturbations}

We know that our model consists of five degrees of freedom, where two are related to gravity, two others to the electromagnetic field and finally one degree of freedom is associated to the scalar field. They are decomposed around a spherically symmetric spacetime as two vector perturbations and three scalar perturbations. As we have seen, vector perturbations propagate at the speed of light. To complete the analysis, we need to obtain the speed propagation of the even perturbations. In that direction, let us consider that the solution is of the form\footnote{We are calculating only the radial speed of propagation.} $\chi \propto e^{i(\omega t - k r)}$. Considering the small scale limit, the dispersion relation obtained from (\ref{eq:Even-Lag-MatrixForm}) can be written as $\det( \omega^2 K_{ij} - k^2 L_{ij}) = 0$. The propagation speed $c_r$ in proper time can be derived by substituting $\omega = \sqrt{AB} c_r k$ into the dispersion relation. Solving for $c_r^2$, we obtain the propagation of the three degrees of freedom:
\begin{equation}
    c_{r1}^2 = c_{r2}^2  = 1 \;, \qquad 
    c_{r3}^2 = \frac{C^2 f_1 \left(3 B {f_1'}^2-2 P_3 f_1\right)}{B \Xi ^2 (2 P_1 -f_1)} \, ,
    \label{vel-highermultipoles}
\end{equation}
where 
\begin{equation}
    P_3 = 2 \Sigma +\frac{\mathcal{M}^2}{\cF}
\end{equation}
with
\begin{align}
    \Sigma = \bar{X} \qty(  f_{2,X} + 2 \bar{X} f_{2,XX})   \,, ~
    \cM = -4 \bar{X} \bar{F} f_{2,FX}                       \,, ~\text{and}~~
    \cF = -2 \bar{F} \qty(  f_{2,F} + 2 \bar{F} f_{2,FF})   \, .
\end{align}
The propagating mode different than the speed of light is related to the scalar field sector. 
Indeed, in the case where the vector field is absent, we have that $P_3 = 2 \Sigma$, and $c_{r3}^2$ reduces to the propagation speed of the scalar field given in \cite{Kobayashi2014}.

Using previous relations, we get\footnote{
    Notice that if we consider a theory for which $f_{2, F} + 2 F f_{2, FF}=0$, viz.
    \begin{align}
        S=\int {\rm d}^4 x \sqrt{-g}\Bigl[F(\phi)R+K(\phi,X)+G(\phi,X)\sqrt{F}\Bigr]
    \end{align}
    one of the degrees of freedom propagates at infinite speed, which could be similar to the Cuscuton \cite{Afshordi:2006ad}, for which this perturbation do not carry any microscopic information.
}
\begin{align}
c_{r3}^2 = 1 + 4\bar{X} f_1 \frac{{f_2}_{,XX}(f_{2,F} + 2 \bar{F} f_{2,FF})-2 \bar{F} {f_2}_{,XF}^2}{(f_{2,F} + 2 \bar{F} f_{2,FF})(3 {f_1}_{,\phi}^2+2 f_1 {f_2}_{,X})} \, .
\label{eq:speed}
\end{align}

It is important to emphasize that these velocities are very weakly constrained by gravitational waves. 
Certainly, between the merger of two compact objects and us, the wave propagates mostly over a FLRW spacetime than a spherically symmetric background. 
But, we should also consider that a model described by our action is just an effective low energy description of some more fundamental theory. For that, we should impose standard conditions such as Lorentz invariance, unitarity, analyticity. 
Even if not proved generically, it was shown in various situations and around different backgrounds that these conditions imply nonsuperluminal propagation (see e.g. \cite{Adams:2006sv,Nicolis:2009qm,Melville:2019wyy}). 
Hereafter, we will not consider this condition but the much more restrictive, possibly more interesting, condition $c_{r_3}^2 = 1$, which translates from eq.(\ref{eq:speed}) into
\begin{equation}
    f_{2, XX} = \frac{2 \bar{F} (f_{2, XF})^2}{f_{2, F} + 2 \bar{F} f_{2, FF}} \, .
    \label{eq:c3_consition}
\end{equation}
We conclude that any action given by the form below will propagate at the speed of light five degrees of freedom
\begin{align}
    S=\int {\rm d}^4 x \sqrt{-g}\Bigl[f_1(\phi)R+f_2(\phi,F)+Xf_3(\phi)\Bigr] \, ,
    \label{eq:speed1}
\end{align}
where $(f_1,f_2,f_3)$ are generic functions.

\subsection{Master equation}

From Lagrangian (\ref{eq:Even-Lag-MatrixForm}), we obtain the equations of motion
\begin{equation}
    L_{ij} \chi_j '' + (L_{ij}' - D_{ij} ) \chi_j' + (M_{ij} + \omega^2 K_{ij} - \frac12 D_{ij}') \chi_j = 0 \, ,
\end{equation}
where we have used $\vec{\chi}(t,r) \to e^{-i\omega t}\vec{\chi}(r)$. 
Making a change of variable $\chi_i(r) \to S_{ij} (r) \Phi_j(r)$, and changing to tortoise coordinate $dr = \sqrt{AB} d r_*$ we get 
\begin{equation}
    \frac{d^2 \Phi_q}{d r_*^2} + \omega^2 \qty(A B S^{-1}_{qp} L^{-1}_{pi} K_{in} S_{nk})\Phi_k + A B S^{-1}_{qp} L^{-1}_{pi} B_{in} S_{nk} \Phi_k = 0
    \label{eq:general-EOM-even}
\end{equation}
where the matrix $S$ is solution of\footnote{This differential equation will have some integration constants which can be left free as soon as $S$ is invertible.} 
\begin{align}
    S'_{ij}  +C_{ik}S_{kj}=0 \, ,
    \label{eq:matrixS}
\end{align}
with
\begin{align}
    C_{ij} = \frac12 L^{-1}_{ik} \qty[L_{kj}' - D_{kj} - \frac12 \qty(\frac{A'}{A} + \frac{B'}{B})L_{kj}] \, ,
\end{align}
and
\begin{align}
    B_{ij}  = L_{ik}(C_{km} C_{mj} - C'_{kj}) - L'_{ik} C_{kj} + D_{ik} C_{kj} + M_{ij} - \frac12 D'_{ij} \, .
\end{align}
We see from eq.(\ref{eq:general-EOM-even}) that not all degrees of freedom propagate at the speed of light. 
But if we consider the condition (\ref{eq:c3_consition}), we have $L_{ij}=ABK_{ij}$, which implies
\begin{equation}
    \frac{d^2 \vec{\Phi}}{d r_*^2} + \omega^2 \vec{\Phi} - \mathbf{V} \vec{\Phi} = 0
    \label{eq:EOM-with-V}
\end{equation}
where $\mathbf{V}$ is the matrix potential, which expression can be read off directly from (\ref{eq:general-EOM-even})
\begin{align}
    \mathbf{V}=-\mathbf{S}^{-1}\mathbf{K}^{-1}\mathbf{B}\, \mathbf{S} \, .
    \label{eq:Vscalar}
\end{align}
Given the matrix $\mathbf{S}$, the potential $\mathbf{V}$ can be easily derived and the stability studied. Unfortunately, in the generic case, we were not able to obtain the matrix $\mathbf{S}$ but we will see in the following sections, various examples where these calculations can be performed easily.

Following the same procedure as in the vector perturbation sector, we define the operator $\mathcal{H}_{ij}=-\delta_{ij}\frac{{\rm d}^2}{{\rm d}r_*}+V_{ij}$, and we obtain
\begin{align}
    (\Phi,\mathcal{H}\Phi) &=\int{\rm d}r_* \bar{\Phi}_i \mathcal{H}_{ij} \Phi_j\nonumber\\
    &=\int{\rm d}r_* \Bigl[\Bigl|\frac{d\Phi_i}{dr_*}\Bigr|^2+V_{ij}\bar\Phi_i\Phi_j\Bigr] \, ,
\end{align}
where we have as usual performed an integration by parts and eliminated the boundary term. Even if a generic stability condition can't be obtained, we can discuss sufficient stability conditions. In fact, a sufficient but not necessary condition of the stability is obtained if the potential is definite positive. On the contrary, if there is a trial function\footnote{Because the lowest eigenvalue $\omega_0$ of the spectrum of $\mathcal{H}$ gives $\forall ~\Phi$, $\omega_0^2\leq (\Phi,\mathcal{H}\Phi)/(\Phi,\Phi)$. Therefore if a trial function $\Phi_0$ is such that $(\Phi_0,\mathcal{H}\Phi_0)<0$, we have $\omega_0^2<0$, for a normalizable trial function.} $\Phi_0$, such that $(\Phi_0,\mathcal{H}\Phi_0)<0$, the spacetime is unstable \cite{Volkov:1994dq}.

In the case\footnote{Notice that this condition includes most, if not all models studied in the literature for generic Einstein-Maxwell-dilaton theories.} where $f_{2,XF}=0$, we found for large $\lambda$ 
\begin{align}
    V_{22}=\lambda \frac{A}{C}\frac{f_{2,F}}{f_{2,F}+2\bar{F} f_{2,FF}}+\mathcal{O}(\lambda^0) \, .
\end{align}
Considering $\Phi_0=(0,\phi_0(r),0)^T$ with $(\phi_0,\phi_0)<\infty$ (a normalizable trial function), we get
\begin{align}
    (\Phi_0,\mathcal{H}\Phi_0) =\int{\rm d}r_* \Bigl[\Bigl|\frac{d\phi_0}{dr_*}\Bigr|^2+\lambda \frac{A}{C}\frac{f_{2,F}}{f_{2,F}+2\bar{F} f_{2,FF}}|\phi_0|^2\Bigr]+\mathcal{O}(\lambda^0) \, .
\end{align}
Therefore we conclude that if 
\begin{align}
    \frac{f_{2,F}}{f_{2,F}+2\bar{F} f_{2,FF}}<0 \, ,
\end{align}
the integral is negative for sufficiently large $\lambda$ which implies the instability.

\subsection{Second order Lagrangian for \texorpdfstring{$\ell = 0$}{l=0}}

Similarly to the vector perturbations, the generic analysis we have performed in the previous section is valid only for higher modes, $\ell\geq 2$. For scalar perturbations, we will see that for $\ell=0$, we have only one perturbation related to the scalar field, the breathing mode, which is the spherically symmetric perturbation and for $\ell=1$ we will have 2 perturbations related to the scalar field and the electromagnetic field. Using the relations derived in Sec.\ref{formalism}, we have, for $\ell=0$, without defining any gauge, $h_{0i}=h_{1i}=0$ and only $K_{00}$ survives in the expression of $h_{ij}$. We can choose the freedom on $\xi_1$ to fix $K_{00}=0$. We still have the freedom on choosing $\xi_0$. Using these relations, we obtain from eq.(\ref{L-evenperturbation}) using $\ell=0$ ($\lambda=0$)
\begin{equation}
\begin{aligned}
\mathcal{L}^{(2)}_{\text{even},~\ell=0} & = 
	a_1 H_0^2 + 
	H_0 \qty[
	a_2 H_2' + a_4 H_2 + a_7 \d\phi + a_9 \d\phi ' + a_{10} \d\phi''  + a_{11} (A_0' - \dot{A}_1) ] \\
    &  
    +H_1 \qty[b_3 \dot{H}_2 + b_5 \dot{\d\phi} + b_6 \dot{\d\phi}'] + c_1 H_2^2  +
    c_5 H_2 \d\phi + c_7 H_2 \d\phi ' + c_8 \dot{H}_2 \dot{\d\phi} \\
    &  
    + c_{11} H_2 (A_0' - \dot{A}_1) + e_1 \dot{\d\phi}^2 + e_2 \d\phi '^2 + e_3 \d\phi^2 + 
    e_5 \d\phi (A_0' - \dot{A}_1) \\ 
    & 
    + e_6 \d\phi ' (A_0' - \dot{A}_1) + d_1 (A_0' - \dot{A}_1)^2 \, .
\end{aligned}
\end{equation}

Variation of the action with respect to $A_0$ and $A_1$ gives us the relations $\p_t S_e(t,r) = \p_r S_e(t,r) = 0$, where $S_e(t,r)$ is defined in (\ref{eq:H-def}); thus, $S_e(t,r) \equiv C_1$ is constant.

Variation wrt $H_0$, and using the previous relation gives
\begin{align}
    \partial_r\Bigl[P(r) \delta \phi + a_{2} H_{2} + a_{10} \delta \phi' - \frac{C_1}{2} \bar{A}_0\Bigr]=0 \, ,
\end{align}
while variation wrt $H_1$ gives
\begin{equation}
    \partial_t\Bigl[\frac{P(r) \delta \phi + a_{2} H_{2} + a_{10} \delta \phi'}{A}\Bigr] = 0 \, .
\end{equation}
Therefore, we conclude that
\begin{align}
    P(r) \delta \phi + a_{2} H_{2} + a_{10} \delta \phi' = \frac{C_1}{2}\bar{A}_0(r)+C_2 \, ,
\end{align}
where $C_2$ is a new integration constant. This equation will be used to eliminate the variable $H_2$. 

We can use the freedom on $\xi_0$, to partially fix the gauge by considering $H_1=0$ which would also gives us an additional freedom viz. $H_0\rightarrow H_0 - f(t)$. Considering now the variation wrt to $H_2$ and fixing this gauge, we obtain an expression for $H_0'$ which can be integrated and the integration constant $g(t)$ eliminated by the remaining gauge freedom. These expressions are not specially illuminating and therefore will not be written but we will demonstrate the effect of this mode on a particular solution. Finally, the equation wrt to $\delta\phi$ gives the equation
\begin{align}
    -K_0\ddot{\delta\phi}+\Bigl(L_0\delta\phi'\Bigr)'+M_0\delta\phi+N_0(r)=0
    \label{scalar0}
\end{align}
where we have defined
\begin{equation}
    K_0 = \frac{C'^2 (2{P_1} - f_1)}{2 \sqrt{AB} C \bar{\phi} '^2} \, .
\end{equation}
From this equation, we can see that the no-ghost condition, $K_0 > 0$, is the same as the one for higher multipoles, i.e., equation (\ref{Stability-Condition}). On the other hand, in the limit of large wavenumber $k$, we get that the velocity of the propagation is
\begin{equation}
c^2_0 = \frac{L_0}{AB K_0} = \frac{C^2 f_1 \left(3 B f_1'^2-2 P_3 f_1\right)}{B \Xi ^2 (2 P_1 -f_1)} \, ,
\end{equation}
which also coincides with $c_{r_3}^2$ given in Eq. (\ref{vel-highermultipoles}). Since only the scalar wave is excited in the monopole perturbation, this result allows us to interpret $c_{r_1}^2$ and $c_{r_2}^2$ as the propagation speed of gravitational and vector perturbations, and $c_{r_3}^2$ as the propagation of the scalar waves.

The eq.(\ref{scalar0}) admits a particular and homogeneous solution. The homogeneous solution will describe the propagation of the spherical scalar wave in the fixed background metric while the particular solution should modify the constants of the metric. For that, let us consider the electric GM-GHS black hole (see section \ref{GMGHS}) with
\begin{align}
    f_1 &= 2\,, ~~f_2=4X+4 e^{-2\phi}F\\
    A&=B=1-\frac{2M}{r}\,,~~C=r\Bigl(r-\frac{q^2}{M}\Bigr)\\
    \bar{A}_0 &=-\frac{q}{r}\,,~~\phi=\frac{1}{2}\log(1-\frac{q^2}{M r})
\end{align}
Using the previous equations, we found that a particular solution of eq.(\ref{scalar0}) is a constant which gives us $(H_0,H_2)$ and therefore
\begin{align}
    h_{00}=\frac{\alpha}{r}\,,\qquad h_{11}=\frac{\alpha}{r(1-2M/r)^2} \, ,
\end{align}
where $\alpha$ is a constant, $\alpha=M(C_1q-4C_2M)/(8M^2-2q^2)$. That gives us $A=B= 1-2M/r-\alpha/r$. Therefore, only the mass has been modified. Also we have
\begin{align}
    A_0'-\dot A_1=\frac{MC_1+4\alpha q}{8Mr^2} \, ,
\end{align}
which shifts the electric charge $F_{01}\rightarrow F_{01}-(MC_1+4\alpha q)/8Mr^2$. In conclusion, we expect that for any spacetime, the particular solution shifts the constants of the spacetime while the homogeneous solution describes a scalar wave propagating in a fixed background.

\subsection{Second order Lagrangian for \texorpdfstring{$\ell = 1$}{l=1}}
Considering now the dipole perturbation, we see that for $\ell=1$ and therefore $Y_1^0\propto \cos\theta$, we have 
\begin{align}
    h_{22} &= (K(t,r)-G(t,r))\cos\theta \, ,\\
    h_{33} &= (K(t,r)-G(t,r))\cos\theta\sin^2\theta \, ,\\
    h_{23} &= 0 \, .
\end{align}
Therefore, the metric perturbation $h_{ij}$ depends on $K$ and $G$ only through the combination $K - G$. We can use one function of the coordinate transformation, $\xi_2$, to set $K - G=0$ ($\xi_3=0$). We also use the transformation $\xi_0$ to define $\beta(t,r)=0$. Thus, we have a remaining degree of freedom that we can use to set $\delta \phi (t,r)= 0$. As we have seen, $\delta\phi\rightarrow \delta\phi-B\bar{\phi}' R$. Using the transformation of coordinate $\xi_1$ sets the scalar field to zero. In this case, the gauge is totally fixed and therefore we can proceed as usual. The Lagrangian for the dipole mode is obtained taking $\l =2$ $(\ell =1)$ in (\ref{L-evenperturbation})
\begin{align}
    \cL_{even,~\ell=1}^{(2)} & = a_1 H_0^2 + H_0 \qty[
    a_2 H_2' + 2 a_3 \alpha' + (a_4 + 2 a_5) H_2 + 2 a_6 \alpha + a_{11} (A_0' - \dot{A}_1)]  \nonumber\\
    & +(b_1 + 2 b_2) H_1^2 + H_1 \qty[
    b_3 \dot{H}_2 + 2 b_4 \dot{\alpha} + b_5 \dot{\d\phi}
    ] + c_1 H_2^2 + 2 c_2 \alpha^2 + 2 c_3 H_2 \alpha + 2 c_4 \dot{\alpha}^2  \nonumber\\
    & + c_{11} H_2 (A_0' - \dot{A}_1) + 2 c_{12} \alpha A_0 +
     d_1 (A_0' - \dot{A}_1)^2 + 2 d_2 A_0^2 + 2 d_3 A_1^2 \, .
    \label{Dipole-EvenLag}
\end{align}
Performing similar calculations to those we did in the section of higher multipoles, after taking $\d\phi=0$, we get a second-order Lagrangian similar to (\ref{EvenL2-Lag}). Again, after redefining the variables, we obtain that the Lagrangian takes the canonical form 
\begin{equation}
\cL_{even,~\ell=1}^{(2)} = 
    K_{ij}^1 \dot{\chi}_i \dot{\chi}_j - L_{ij}^1 \chi'_i \chi'_j  + D_{ij}^1 \chi'_i \chi_j  + M_{ij}^1 \chi_i \chi_j \, .
    \label{Lag_Chi}
\end{equation} 
In this case, $\chi$ is a two-component vector representing the two propagating degrees of freedom, $K^1$, $L^1$ and $M^1$ are $2\times 2$ symmetric matrices and $D^1$ is antisymmetric. 
The no-ghost conditions, positivity of the matrix $K_{ij}^1$, are the same as in the higher multipole case, i.e., equations (\ref{K11-Cond}) and (\ref{Inner-DetCond}), which are reduced to the stability condition (\ref{Stability-Condition}) after taking $\l=2$. 
Also, from this Lagrangian, we obtain that the propagation speeds along the radial direction are given by
\begin{equation}
c_1^2 = 1 \, , \qquad\qquad
c_s^2 = \frac{C^2 f_1 \left(3 B f_1'^2-2 P_3 f_1\right)}{B \Xi ^2 (2 P_1 -f_1)} \, ,
\end{equation}
which again coincide with the propagation speed in the case of higher multipoles Eq. (\ref{vel-highermultipoles}). 
We conclude that the degree of freedom associated to the scalar field travels at the same speed, regardless of the multipole order.

\section{Application to specific models}

\subsection{Reissner–Nordström BH}
As a first example, let us consider the Reissner–Nordström BH, which is given by setting $f_1 = 2$ and $f_2 = 4 F$. We have four DOF, two in each type of perturbations, traveling at the speed of light.
The metric functions are
\begin{equation}
    A(r) = B(r) = 1 - 2M/r + q^2/r^2 \,, \qquad C(r) = r^2  \, .
\end{equation}
Because $f_{2,F}=4>0$, we conclude that the vector perturbations are stable. For completeness, let us write the perturbation potential for this sector
\begin{align}
    V_{11} &=\frac{\left(r (r-2 M)+q^2\right) \left(r (\lambda  r-6 M)+4 q^2\right)}{r^6} \, ,\\
    V_{22} &=\frac{\left(r (r-2 M)+q^2\right) \left(4 q^2+\lambda  r^2\right)}{r^6} \, ,\\
    V_{12} &=\frac{2  q \left(r (r-2 M)+q^2\right)}{r^5}\sqrt{\lambda -2} \, ,
\end{align}
which corresponds exactly to the potential derived\footnote{Our perturbations $(V_g,V_e)$ correspond respectively to $(\hat\pi_g,\hat\pi_f)$ of \cite{Moncrief74-a}.} originally in \cite{Moncrief74-a}.

Considering the scalar perturbation sector, we need first to eliminate one row and one column of our matrices because of the absence of the scalar field. Then, we need to solve the eq. (\ref{eq:matrixS}) which gives
\begin{align}
    S = 
    \begin{pmatrix}
        \frac{ 6 M r-4 q^2+(\lambda-2)  r^2}{r^2}c_1 & \frac{ 6 M r-4 q^2+(\lambda-2)  r^2}{r^2}c_2 \\
        \frac{8 c_1 q}{r}+c_3 & \frac{8 c_2 q}{r}+c_4 
    \end{pmatrix}
\end{align}
where $(c_1,c_2,c_3,c_4)$ are constants of integration, which should be chosen freely as soon as $c_1 c_4-c_2c_3\neq 0$ in order for $S$ to be invertible. 
Finally, the calculation of the scalar potential is trivial and we found that our potential $V$ is similar to the original potential\footnote{Where our variables $(S_g,S_e)$ correspond respectively to $(Q,H)$ of \cite{Moncrief74-b}} derived in \cite{Moncrief74-b} with a change of a constant basis matrix. Both matrices describe the same problem. 
In fact, if we have
\begin{align}
 - \frac{\partial^2\mathbf{\Psi}}{\partial t^2} + \frac{\partial^2\mathbf{\Psi}}{\partial r_*^2}-\mathbf{V}\mathbf{\Psi}=0 \, ,
\end{align}
we can always define a transformation $\mathbf{\Phi}=P\mathbf{\Psi}$ and the similar potential $\mathbf{W}=P\mathbf{V}P^{-1}$ such that
\begin{align}
 - \frac{\partial^2\mathbf{\Phi}}{\partial t^2} + \frac{\partial^2\mathbf{\Phi}}{\partial r_*^2}-\mathbf{W}\mathbf{\Phi}=0
\end{align}
if $P$ is a constant matrix.

From the potential, the stability can be easily obtained because of the positivity of the matrix. 

\subsection{Nonlinear electrodynamics}

In this section, we generalize the previous section by considering nonlinear electrodynamics (NED) black holes, given by the action
\begin{equation}
    S = \int d^4x \sqrt{-g}\qty[\frac{R}{4} + \cL(F)] \, ,
\end{equation}
where $\cL$ is an arbitrary function of $F=-F_{\mu\nu}F^{\mu\nu}/4$. From the background equations, it is easy to find that
\begin{align}
    \frac{A'}{A}-\frac{B'}{B}+\frac{C'}{C}-2\frac{C''}{C'}=0 \, ,
\end{align}
which can be easily integrated to
\begin{align}
    \frac{AC}{BC'^2}= \alpha \, ,
    \label{eq:alpha}
\end{align}
where $\alpha$ is a constant of integration. 
In the case where we would be interested only to the background solution, we could always reparametrize our time coordinate such that $\alpha=1$ but because the perturbations are time dependent we will keep that constant. 

The vector perturbations are stable if the condition (\ref{VecStability}) is satisfied ($f_1(\phi)=1/2>0$), viz. 
\begin{align}
    \cL_{,F}>0 \, .
\end{align}
For completeness of the vector perturbations, the potential (\ref{Va_potential},\ref{Vb_potential},\ref{Vc_potential}) is given by
\begin{align}
    V_{11} & = (\lambda-2)\frac{A}{C}+\frac{ABC'}{2C}\Bigl(\frac{C'}{C}-\frac{A'}{A}\Bigr)  \, ,\\
    V_{12}  &= \frac{2 A}{C}\sqrt{\frac{Q^2(\lambda -2)}{C{\cL_{,F}}}}  \, ,\\
    V_{22} & = \frac{A}{C}\Bigl(\lambda+\frac{4Q^2}{C{\cL_{,F}}}\Bigr)+\frac{AB{\cL_{,F}'}}{4{\cL_{,F}}}\Bigl(\frac{A'}{A}+\frac{B'}{B}+2\frac{{\cL_{,F}}''}{{\cL_{,F}}'}-\frac{{\cL_{,F}}'}{{\cL_{,F}}}\Bigr) \, .
\end{align}
In \cite{Moreno:2002gg}, the perturbations for this model were studied in a gauge invariant formalism. We find exactly the same potential when $C(r) = r^2$.

Focusing on the even-parity sector, since we have only two degrees of freedom, the stability conditions reduce to (for simplicity of the expression, we have eliminated the obvious positive factors)
\begin{equation}
    K_{11} \propto \lambda-2+4 Q^2 C^{-1}\cL_{,F}^{-1} >0 \, , \qquad
    \det(K) \propto (\lambda -2) \mathcal{L}_{,F}^{-1} >0 \, ,
\end{equation}
giving the same condition as the odd-parity sector, namely $\mathcal{L}_{,F}>0$. Following with the even-parity sector, we compute the matrix $S$ (\ref{eq:matrixS})
\begin{align}
    S=
    \begin{pmatrix}
     \frac{CA'-AC'+2 \alpha \lambda C'}{C'} c_1 & \frac{CA'-AC'+2 \alpha \lambda C'}{C'}c_2 \\
     -\frac{8 \sqrt{\alpha} Q}{\sqrt{C}}c_1+\frac{\sqrt{C'}}{C^{3/4}\sqrt{\bar{A}_0'}}c_3 & -\frac{8 \sqrt{\alpha} Q}{\sqrt{C}}c_2 +\frac{\sqrt{C'}}{C^{3/4}\sqrt{\bar{A}_0'}}c_4
\end{pmatrix}
\end{align}
where as in the previous case, the constants $(c_1,c_2,c_3,c_4)$ can be freely defined as soon as $c_1 c_4-c_2c_3\neq 0$. The expression of the potential in the generic case is long and not necessarily instructive. It can be easily obtained from the eq.(\ref{eq:Vscalar}). It is anyway interesting to derive the most used expression in the literature, namely $A(r)=B(r)$ and $C(r)=r^2$ which implies from eq.(\ref{eq:alpha}), $\alpha=1/4$. We will take without any restriction $c_2=c_3=0$ and $c_4=\sqrt{2Q(\lambda-2)}c_1$. We get
\begin{align}
    V_{11} & = \frac{A\qty[\lambda  (\lambda -2) + r A' (a+2) - 2 A (\lambda -2)]}{r^2 (\lambda +a)} +\frac{2 A^2 (\lambda -2)\mathcal{B}}{r^2(\lambda + a)^2}  \, ,\\
    V_{12} & = \cV_{21} =\frac{2 A \sqrt{Q^2 (\lambda -2)}}{r^3 (\lambda + a) \sqrt{\cL_{,F}}} \left(\lambda -a + 2 A \kappa -4 A+\frac{2 A \mathcal{B}}{\lambda + a} \right)  \, , \\
    V_{22} & = A \left(\frac{\kappa\lambda }{r^2} + \frac{4 Q^2 \left(\lambda -2 A+4 \kappa A - r A'\right)}{\cL_{,F} \left(\lambda + a\right) r^4} + \frac{8 Q^2 A \mathcal{B}}{r^4 \cL_{,F} \left(\lambda + a\right)^2}+ \cL_{,F}^{1/2} \left[A (\cL_{,F}^{-1/2})'\right]'\right) \, ,
\end{align}
where we have defined $a(r) = r A' - 2 A$, $\mathcal{B} = \lambda - 2 + 4 Q^2 / r^2 \cL_{,F}$ and $\kappa=-r \bar{A}_0''/(2\bar{A}_0')=\cL_{,F}/(\cL_{,F}+2\bar{F} \cL_{,FF})$. This potential is the same as in \cite{Moreno:2002gg}. From which we can easily derive the stability condition for a specific model or calculate the QNMs. 

It is important to notice that the stability will depend only on the metric potential $A$ and its derivatives. In fact, using the equations of motion, we can eliminate the electric potential $\bar{A}_0$ and the Lagrangian $\cL$, reducing the potential matrix to
\begin{align}
    V_{11} &= \frac{A}{r^2} \Bigl[r A'-\lambda +2+\frac{2 (\lambda -2) (\lambda -2 A)}{r A'-2 A+\lambda }+A\frac{2 (\lambda -2) \left(r^2 A''-2 A+\lambda \right)}{\left(r A'-2 A+\lambda \right)^2}\Bigr]  \, ,
    \label{eq:V11}\\
    V_{12} &=\sqrt{\lambda-2}\frac{A}{r^2}\sqrt{2-2A+r^2 A''}\Bigl[\frac{\lambda-4A-rA'}{\lambda-2A+rA'}+2A\frac{\lambda-2A+r^2A''}{(\lambda-2A+rA')^2}\nonumber\\
    &\qquad +A\frac{4-4A+2rA'-r^3A'''}{(\lambda-2A+rA')(2-2A+r^2A'')}\Bigr]  \, , \label{eq:V12}\\
    V_{22} &=\frac{A}{r^2}\Bigl[
    2 r A'+7A-2-\lambda-r^2A''
    +2 \frac{8 A^2 + \lambda (2 + r^2 A'') - 
  A (4 + 4 \lambda + 4 r^2 A'' + r^3 A''')}{\lambda - 2 A + 
  r A'}\nonumber\\
  &\qquad +\frac{2 A (2 - 2 A + r^2 A'') (\lambda - 
   2 A + r^2 A'')}{(\lambda - 2 A + r A')^2}
  -\frac{A (4 - 4 A + 2 r A' - r^3 A''')^2}{4 (2 - 2 A + r^2 A'')^2}\nonumber\\
  &\qquad +\frac{(\lambda - r A') (4 + 2 r A' -  r^3 A''') +  A (-4 \lambda + 4 r A' + 4 r^3 A''' + r^4 A'''')}{2 (2 - 2 A + r^2 A'')}
    \Bigr]  \, .
    \label{eq:V22}
\end{align}

From which we conclude that different theories (Lagrangians) having the same black hole solution will have the same stability. In that direction, we found that considering $A=B=1-2M/r+q^2/r^2$, the stability potential reduces to the Moncrief potential\footnote{Our potential $V$ can be written as $V=PWP^{-1}$ with $W$ the Moncrief potential and $P=\begin{pmatrix}
 0 & 1 \\
 -1 & 0 
\end{pmatrix}
$, therefore they are similar.} which implies the stability of the Reissner-Nordstr{\"o}m metric independently of the nonlinear electrodynamics theory considered. 

\subsubsection{Bardeen black hole}

As an application, let us consider one of the first singularity free black hole proposed by Bardeen \cite{Bardeen}. This metric was shown to be an exact solution of Einstein equations with a nonlinear magnetic monopole \cite{Ayon-Beato:2000mjt}. It is described by the metric
\begin{align}
    A=B=1-\frac{2Mr^2}{(r^2+q^2)^{3/2}}
\end{align}
and $C=r^2$. Because we have focused in this paper on electric source, we know from the FP duality\footnote{The Lagrangian $\mathcal{L}(F)$ can also be written in terms of the field $P=P_{\mu\nu}P^{\mu\nu}/4$ where $P_{\mu\nu}=F_{\mu\nu}\mathcal{L}_{,F}$ \cite{Salazar:1987ap}. This is the so-called P framework. The equations in the P framework are equivalent to the equations in the F framework by performing a transformation while the metric remains unchanged \cite{Bronnikov:2000vy}. Therefore, the FP duality connects theories with different Lagrangians but similar metric. A purely magnetic solution in the P framework will correspond to a purely electric solution in the F framework or vice versa. This transformation exists \cite{Moreno:2002gg} if $\mathcal{L}_{,F}(\mathcal{L}_{,F}+2F \mathcal{L}_{,FF})\neq 0$.}, that a solution can be generated by two different Lagrangians, one theory with a magnetic field and the other with an electric field\footnote{It would be interesting to see if a magnetically charged BH follows the same stability condition as the electrically charged BH. Indeed, we found previously that in the electrically charged case, the stability condition is independent of the Lagrangian.}. It is easy to find that a Lagrangian with electric field generating the Bardeen spacetime is
\begin{align}
    \bar{A}_0(r) & = \alpha \frac{r^5}{( r^2+ q^2)^{5/2}} \, ,\\
    \cL (r) &= \frac{3 M q^2 (3 r^2-2 q^2)}{2(r^2+q^2)^{7/2}} \, ,
\end{align}
where $\alpha$ is a constant. From which, we could obtain $\cL(F)$ but because the expression is numerical and not analytical, we will just mention that such Lagrangian exists\footnote{It was shown in \cite{Bronnikov:2000vy} that a Lagrangian having a Maxwell asymptotic "\textit{does not admit a static, spherically symmetric solution with a regular center and a nonzero electric charge}", which seems to contradict our result. In fact, the Lagrangian we found is a multivalued function, therefore we would need different Lagrangians for different ranges of the radial coordinate $r$. But it is single-valued if we restrict our analysis to the exterior region, which is the interesting area for our analysis. Of course, these models can't be continued until the singularity and therefore loose their interest as singular free models.}. We can see from the electric potential, that we will not recover the Coulomb potential at large distances, which could also be seen from the metric that reduces to $1-2M/r+3Mq^2/r^3$ for large $r$. Therefore, even at large distances, the black hole is not similar to Reissner-Nordstr\"om spacetime. But interestingly, long distance quantum corrections to the Schwarzschild black hole goes like $1/r^3$ \cite{Duff:1974ud,Bjerrum-Bohr:2002fji} which is similar to the Bardeen solution if we identify $q^2$ to $62\hbar/45\pi$

Looking now to the stability of this black hole, we found that odd-parity perturbations are stable if $f_{2,F}=3M/\alpha r^2$ is positive which implies the trivially satisfied condition {$M\alpha~>~0$}. 

Considering finally the even parity sector, we have analysed the stability numerically from the potential (\ref{eq:V11},\ref{eq:V12},\ref{eq:V22}). For that, we redefined the variables $r\rightarrow \bar{r}q$, $M\rightarrow \bar{M}q$ such that $\bar{M}$ remains the only free parameter. Considering the normalized mass in the range $0<\bar{M}<10$ and for $\ell={2,...,10}$, we found that the potential is definite positive and therefore the black hole is stable in this range.

\subsubsection{Hayward black hole}

An other interesting singular free black hole is defined as \cite{Hayward:2005gi}
\begin{align}
    A=B=1-\frac{2 M r^2}{r^3+2Ml^2} \, .
\end{align}
This black hole is known as the Hayward black hole. It reproduces the Schwarzschild spacetime at large distances with corrections $\mathcal{O}(1/r^4)$ which are not similar to loop quantum contribution as noticed in the Bardeen solution. But similarly, the black hole is regular at the origin and has a de Sitter core, $A\simeq 1-r^2/l^2$.

As in the previous case, the solution can be derived from various theories. We will consider nonlinear electrodynamics. It is easy to find that a Lagrangian exists, numerically, which gives this metric
\begin{align}
    &\bar{A}_0(r)=-\alpha \frac{6l^2M^2(r^3+M l^2)}{(r^3+2Ml^2)^2}  \, ,    \\
    &\mathcal{L}(r)=\frac{12 M^2 l^2(r^3-Ml^2)}{(r^3+2 M l^2)^3}    \, ,
\end{align}
where $\alpha$ is a constant.

Within this theory, we would like to know if the black hole is stable. The odd-parity sector is trivially stable because $f_{2,F}=(r^3+2Ml^2)^3/18\alpha^2M^2l^2r^7>0$

As for the Bardeen black hole, the even-parity perturbations will be studied numerically. Interestingly this spacetime is invariant under a scaling transformation \cite{Frolov:2016pav}. Therefore, we can renormalize our variables, $r\rightarrow l \sqrt{M} r$, $M\rightarrow M^{3/2} l$, which reduces the solution to only one parameter. We have checked the stability for  $0<(M/l)^{2/3}\leq 10$ and $2\leq l\leq 10$. We found that the potential is definite positive and therefore the black hole is stable in this range.

\subsubsection{Born-Infeld gravity}

Considering open superstring theory, loop contributions lead to a Lagrangian of the Born-Infeld type
\begin{align}
    \mathcal{L}=\frac{4}{b^2}\Bigl[\sqrt{-g}-\sqrt{|\text{det}(g_{\mu\nu}+bF_{\mu\nu}|}\Bigr] \, ,
\end{align}
where $b$ is related to the tension of the D-brane \cite{Tseytlin:1999dj}. In four dimensions, the determinant can be expanded into \cite{Cederwall:1996uu}
\begin{align}
    \mathcal{L}=\frac{4}{b^2}\Bigl[1-\sqrt{1+\frac{b^2}{2}F^2-\frac{b^4}{16}(F\star F)^2}\Bigr] \, ,
\end{align}
where $\star F$ is the dual of the electromagnetic tensor, $(\star F)_{\mu\nu}=\frac{1}{2}\epsilon_{\mu\nu\rho\sigma}F^{\rho\sigma}$. Focusing from now on an electric source and adding gravity, the action becomes
\begin{align}
    S=\int {\rm d}^4 x\sqrt{-g}\Bigl[R+\frac{4}{b^2}\Bigl(1-\sqrt{1+\frac{b^2}{2}F_{\mu\nu}F^{\mu\nu}}\Bigr)\Bigr] \, .
\end{align}
The static spherically symmetric electrically charged black hole was obtained in \cite{Fernando:2004pc}
\begin{align}
    &A=B=1-\frac{2M}{r}+\frac{2}{3b^2}r^2+\frac{2}{\beta^2 r} g(r)\\
    &C=r^2\\
    &\bar{A}_0=-Q\int\frac{{\rm d}r}{g'(r)}
\end{align}
where $g(r)=-\int \sqrt{r^4+b^2 Q^2}{\rm d}r$ which can be expressed in terms of hypergeometric functions.

This solution has interesting features such as recovering the Reissner-Nordstr\"om solution at large distances, with highly suppressed corrections
\begin{align}
    A=B= 1-\frac{2m}{r}+\frac{Q^2}{r^2}-\frac{Q^4 b^2}{20 r^6}+\mathcal{O}\Bigl(\frac{1}{r^{10}}\Bigr) \, ,
\end{align}
where $m=M+\frac{\Gamma(\frac{1}{4})^2 Q^{3/2}}{6\sqrt{\pi b}}$. Also notice that around the singularity, at $r=0$, the metric takes the form $A\simeq -2M/r$ but because $M$ is not the ADM mass, it can take any sign causing a time-like or space-like singularity according to the sign of $m-\Gamma(\frac{1}{4})^2 Q^{3/2}/6\sqrt{\pi b}$. But contrary to \cite{Fernando:2005bc}, we conclude that the singularity is unavoidable, even if $M=0$, because the curvature scalar is $R=4Q/b r^2$ around the singularity.

It is trivial to see that odd-parity perturbations are stable because $f_1=2>0$ and $f_{2,F}=4/\sqrt{1+b^2 F/2}>0$. Turning now, to the even-parity perturbations, we have checked for a large range of parameters and found that the eigenvalues of the matrix $V$ (\ref{eq:V11},\ref{eq:V12},\ref{eq:V22}) are positive outside the horizon, from which we can conclude the stability of this black hole. 

\subsection{Scalar-tensor theory}
After studying spacetimes with the presence of an electric charge, we will consider another particular case of our analysis for which only the scalar field is present, viz.
\begin{align}
    \mathcal{S}&=\int{\rm d}^4x\sqrt{-g}\Bigl[f_1(\phi)\frac{R}{2}+f_2(\phi,X)\Bigr]\\
    &=\int{\rm d}^4x\sqrt{-g}\Bigl[F(\phi)\frac{R}{2}+K(\phi,X)\Bigr]
    \label{Kfield}
\end{align}
where we have redefined the functions of the Lagrangian in a more standard notations of a K-field non-minimally coupled to gravity also known as K-inflation \cite{Armendariz-Picon:1999hyi} or K-essence \cite{Armendariz-Picon:2000nqq}.

Since Nordstr\"om and his attempt to write a scalar theory of gravity \cite{Ravndal:2004ym}, scalar fields remained a very dynamical field of research from phenomenological models to more advanced and motivated theories. It has been motivated as early as Kaluza and Klein where a scalar field appears in the compactification of a fifth dimension. Similarly, string theory behaves in the low-energy limit as general relativity coupled to a scalar field, the dilaton, and a totally
antisymmetric field strength $H_{\mu\nu\rho}$ \cite{Green:1987sp}. Also, we should mention that the dynamics of the tachyon field near the minimum of its potential is given by \cite{Garousi:2000tr,Sen:2002in} $K(\phi,X)=-F(\phi)\sqrt{1+(\partial_\mu \phi)^2}$, or the attempts to describe dark energy by the ghost condensation scenario \cite{Arkani-Hamed:2003pdi} or inflation \cite{Arkani-Hamed:2003juy}. It would be impossible to list all the different models with a scalar field, but we should mention that these models since Brans and Dicke have been largely used as a model testing of general relativity \cite{Will:2014kxa}. 

As in the previous cases, the odd-parity sector is trivial. Indeed, we only need to satisfy the condition $F(\phi)>0$ outside the black hole. For example, any theory expressed in the string frame \cite{Veneziano:1997pz} where $F(\phi)=e^{-\phi}$ would be trivially stable under odd-parity perturbations.

A generic sufficient condition for the stability of even-parity perturbations has not been found in closed form but we will work out particular examples. This analysis can be easily generalized to any BH within this class of theories with the {\scshape Mathematica}\textsuperscript{\textregistered} notebook file provided in \cite{mathematica}. 

Among all these theories, it was shown that Horndeski theories and Lovelock are weakly hyperbolic but might fail to be strongly hyperbolic \cite{Papallo:2017ddx,Papallo:2017qvl,Kovacs:2020ywu}. The only sub-class which is strongly hyperbolic are the K-field models defined in eq. (\ref{Kfield}). Unfortunately for most of these models, a non trivial black hole seems to be impossible. For example, it was shown that if the model is shift symmetric, viz. $f_1=1$ and $f_2=f_2(X)$, the only solution is Schwarzschild \cite{Hui:2012qt}. For that, it was assumed that the area of constant r-spheres should neither be infinite nor zero at the horizon. Violating this condition, the so-called cold black holes (with zero surface gravity) can be constructed \cite{Bronnikov:2006qj}. For that, we need to violate the null energy condition by considering $f_1=1$ and $f_2=-X$. We see that the scalar field has the "wrong" sign for the kinetic term. The solution is
\begin{align}
    {\rm d}s^2 = -\Bigl(1-\frac{2k}{r}\Bigr)^{m/k}{\rm d}t^2+\frac{{\rm d}r^2}{\Bigl(1-\frac{2k}{r}\Bigr)^{m/k}}+r^2\Bigl(1-\frac{2k}{r}\Bigr)^{1-m/k} {\rm d}\Omega^2
\end{align}
with 
\begin{align}
    \phi(r)=-\frac{\sqrt{\frac{m^2}{k^2}-1}}{\sqrt{2}}\ln\Bigl(1-\frac{2k}{r}\Bigr) \, ,
\end{align}
where $m$ is the ADM mass and $k$ is related to the scalar charge. Indeed, we have at infinity $\phi\simeq \frac{\sqrt{2m^2-2k^2}}{r}$.

Even if very interesting, this black hole violates the no-ghost condition (\ref{eq:Sylvester-criterion}). In fact, it is easy to show that $\det(K)<0$. Maybe a stable black hole could be constructed in that direction by considering a non-linear kinetic term. We should here mention that various papers on wormholes consider a ghost-like scalar field. While performing a stability analysis at the equation level and not the action level, we could reach the erroneous conclusion of a normal behaviour of that solution. But because the Hamiltonian is unbounded from below, any coupling to matter such as a detector or any nonlinear regime of the theory would couple to the negative energy modes and render the theory unstable. These theories are pathological.

Another way to escape the no-go theorem previously mentioned is to brake the shift invariance and assume a  potential. We know that important no-hair theorems restrict the existence of non-trivial black holes, see e.g. \cite{Bekenstein:1972ny,Bekenstein:1995un,Sudarsky:1995zg,Herdeiro:2015waa}. One particular solution is given by a scalar field non-minimally coupled to the scalar curvature, the so-called BBMB black hole \cite{Bocharova:1970,Bekenstein:1974sf,Bekenstein:1975ts}. The action is given by $f_1(\phi)=\frac{1}{2}-\frac{\phi^2}{6}$ and $f_2(\phi,X)=X$ while the line element is given by
\begin{align}
      {\rm d}s^2 = -\Bigl(1-\frac{M}{r}\Bigr)^{2}{\rm d}t^2+\frac{{\rm d}r^2}{\Bigl(1-\frac{M}{r}\Bigr)^{2}}+r^2 {\rm d}\Omega^2
\end{align}
and $\phi(r)=\sqrt{3} M/(r-M)$, where $M$ is the ADM mass. The horizon is located at $r=M$ where the scalar field diverges but with a regular geometry of the horizon. It is also a cold black hole which connects smoothly to the Minkowsky space $M=0$ but never to the Schwarzschild spacetime. In fact, the black hole has the same causal structure than the extremal Reissner-Nordstrom black hole. We conclude easily that the black hole is unstable because it violated the condition $f_1(\phi)>0$ in the region $M<r<2M$, which is consistent with the radial perturbations performed in \cite{Bronnikov:1978mx}.

\subsection{BH in Einstein-Maxwell-dilaton theory}
\label{GMGHS}
Finally, we consider the presence of all fields, namely the metric, the scalar and vector field. In this context, an interesting solution was derived in Einstein-Maxwell-dilaton (EMd) theory described by the Lagrangian \cite{Gibbons88, Garfinkle91} 
\begin{align}
S&=\int d^4x \, \sqrt{-g} \left(R -2 \partial_\mu\phi\partial^\mu\phi  -e^{-2a\phi} F_{\mu\nu}F^{\mu\nu}\right) \, ,
\label{eq:EMd-action}
\end{align}
where $a$ is the dilaton coupling constant. The model appears as a low energy limit of heterotic string theory for $a=1$ and as the dimensionally reduced five dimensional vacuum Einstein action in the Einstein frame, for $a=\sqrt{3}$. The static and spherically symmetric black hole solution in the case where the Maxwell field is electric is given by
\begin{align}
    & A=B=\Bigl(1-\frac{r_+}{r}\Bigr)\Bigl(1-\frac{r_-}{r}\Bigr)^{\frac{1-a^2}{1+a^2}} \, ,\\
    & C=r^2 \Bigl(1-\frac{r_-}{r}\Bigr)^{\frac{2a^2}{1+a^2}} \, ,\\
    & \bar{A}_0=-\frac{\sqrt{r_+ r_-}}{\sqrt{1+a^2}~r} \, ,\\
    & \bar{\phi}=\frac{a}{1+a^2}\log\Bigl(1-\frac{r_-}{r}\Bigr) \, .
\end{align}
Here $r_+$ is the position of the event horizon and $r_-$ corresponds to a curvature singularity $R_{\mu\nu\rho\sigma} R^{\mu\nu\rho\sigma}\propto (r-r_-)^{-2(1+3a^2)(1+a^2)}$ except for $a=0$ where the solution reduces to Reissner-Nordstr\"om metric. The two parameters $(r_+,r_-)$ are related to the ADM mass $M$ and the electric charge $Q$ by \cite{Nozawa:2018kfk}
\begin{align}
    r_\pm = \frac{1+a^2}{1\pm a^2}\Bigl(M\pm \sqrt{M^2-(1-a^2)Q^2}\Bigr) \, .
\end{align}
Given the background solution, we can perform the analysis of perturbation theory. As a first step, we focus on odd-parity perturbations. 
The no-ghost conditions (\ref{VecStability}) are trivially satisfied since $f_1(\phi)=2>0$ and $f_{2, F} (\phi, X, F) = 4 e^{-2a\phi} > 0$. We conclude that the odd-parity perturbations are stable.
From Eqs.(\ref{Va_potential}) -- (\ref{Vc_potential}), we get that the components of the potential associated with this type of perturbations are
\begin{align}
    V_{11} & = V_{22} + \frac{\mathcal{T}}{r} \left[\frac{\left(a^2-3\right) {r_-}}{1+a^2}-3 {r_+}\right]
    \, , 
    \\
    V_{12} & = \frac{2\mathcal{T}}{r} \sqrt{\frac{(\lambda -2) {r_-} r_+}{a^2+1}} \, ,\\
    V_{22} & = \mathcal{T} \qty[ \lambda + \frac{r_-}{\left(a^2+1\right)^2 r} \left(\frac{a^4 (2 r+{r_-}-3 {r_+})}{r-{r_-}} + 2 a^2 + \frac{a^2 r_+}{r} + \frac{4 {r_+}}{r}\right)] \, ,
\end{align}
where we have defined
\begin{equation}
    \mathcal{T} = \frac{A(r)}{(r-r_-)^2} \left(1-\frac{r_-}{r}\right)^{2/(a^2+1)}  \, .
\end{equation}

It can be checked that the potential matrix has constant eigenvalues; therefore, we can diagonalize this matrix to decouple the system of wave-like equations. After this diagonalization, the potentials are
\begin{equation}
    \begin{aligned}
    V_\pm^{\mbox{\tiny EMd}} = \frac{\mathcal{T}}{r}\left[ \lambda r + 
    \frac{3 a^4 (r-r_+) r_-^2 }{r (a^2+1)^2 (r-r_-)} + 
    \frac{\left(4-3 a^2\right) r_- r_+}{(a^2+1) r} + 
    \frac{\left(5 a^2-3\right) r_-}{2 \left(a^2+1\right)} - \frac{3 r_+}{2} \right. \\
    \pm \left.\frac12 \sqrt{9 r_+^2+\frac{\left(a^2-3\right)^2 r_-^2}{\left(a^2+1\right)^2}-\frac{2 \left(7+3 a^2-8 \lambda \right) r_- r_+}{a^2+1}}\right] \, .
\end{aligned}
\end{equation}
As a specific case, setting $a =1$, we recover the potential computed in \cite{Ferrari01}.

\begin{figure}[H]
\centering
\includegraphics[scale=.5]{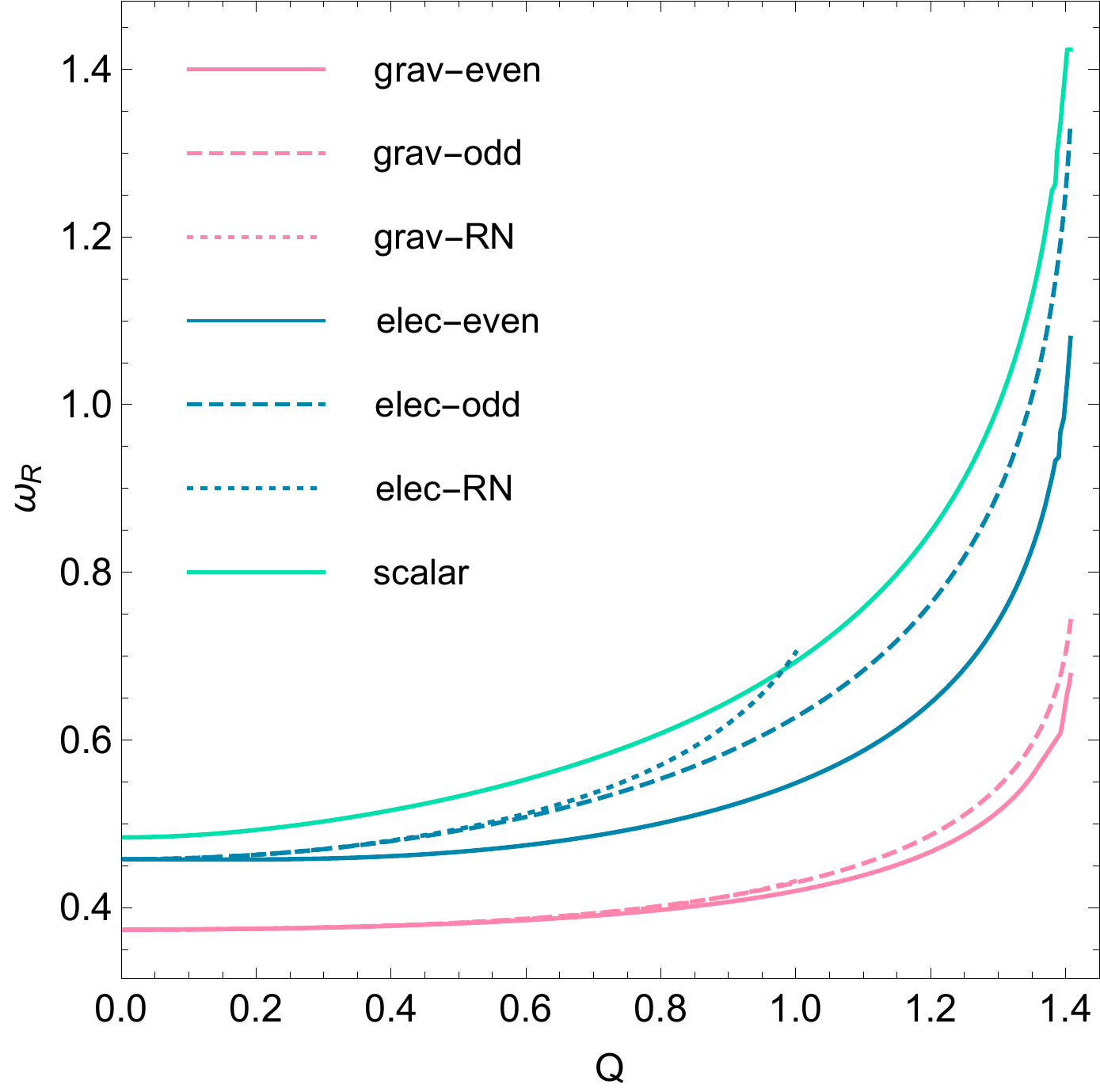}~~~~
\includegraphics[scale=.5]{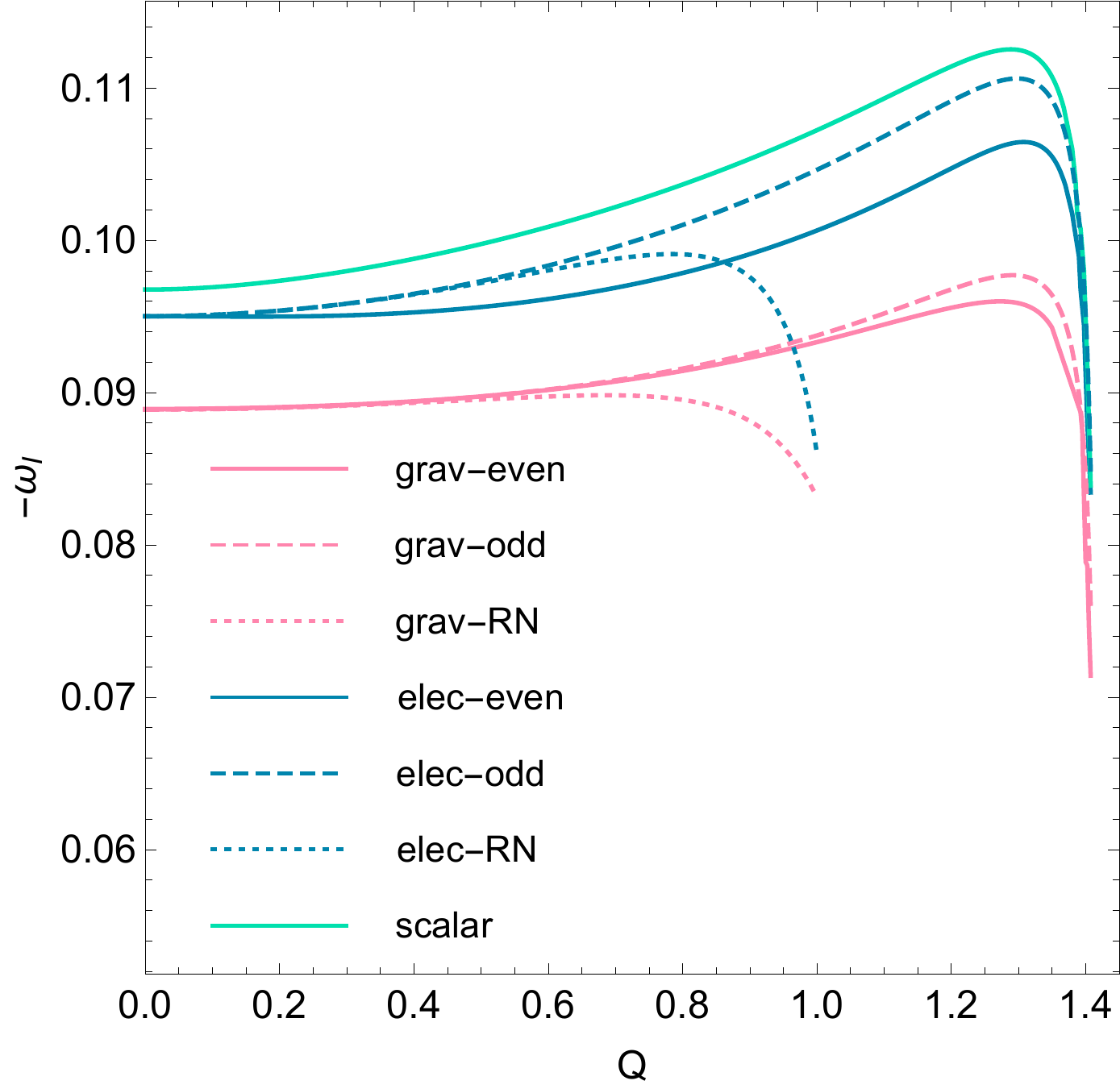}
\caption{Fundamental mode (lowest QNMs) for the EMd black hole as a function of the charge $Q$ in units $M=1$ and for $\ell=2$. In pink, we have the modes associated to the gravitational sector, which reduce to the Regge-Wheeler and Zerilli potentials. In blue, the QNM associated to the electromagnetic sector and finally in blue the mode associated to the scalar field. These perturbations are compared to the Reissner-Nordstrom black hole in dotted line. The QNM have been calculated with a step size of $0.0025$ for the charge. }
\label{QNM-GHS}
\end{figure}

\begin{figure}[H]
\centering
\includegraphics[scale=.5]{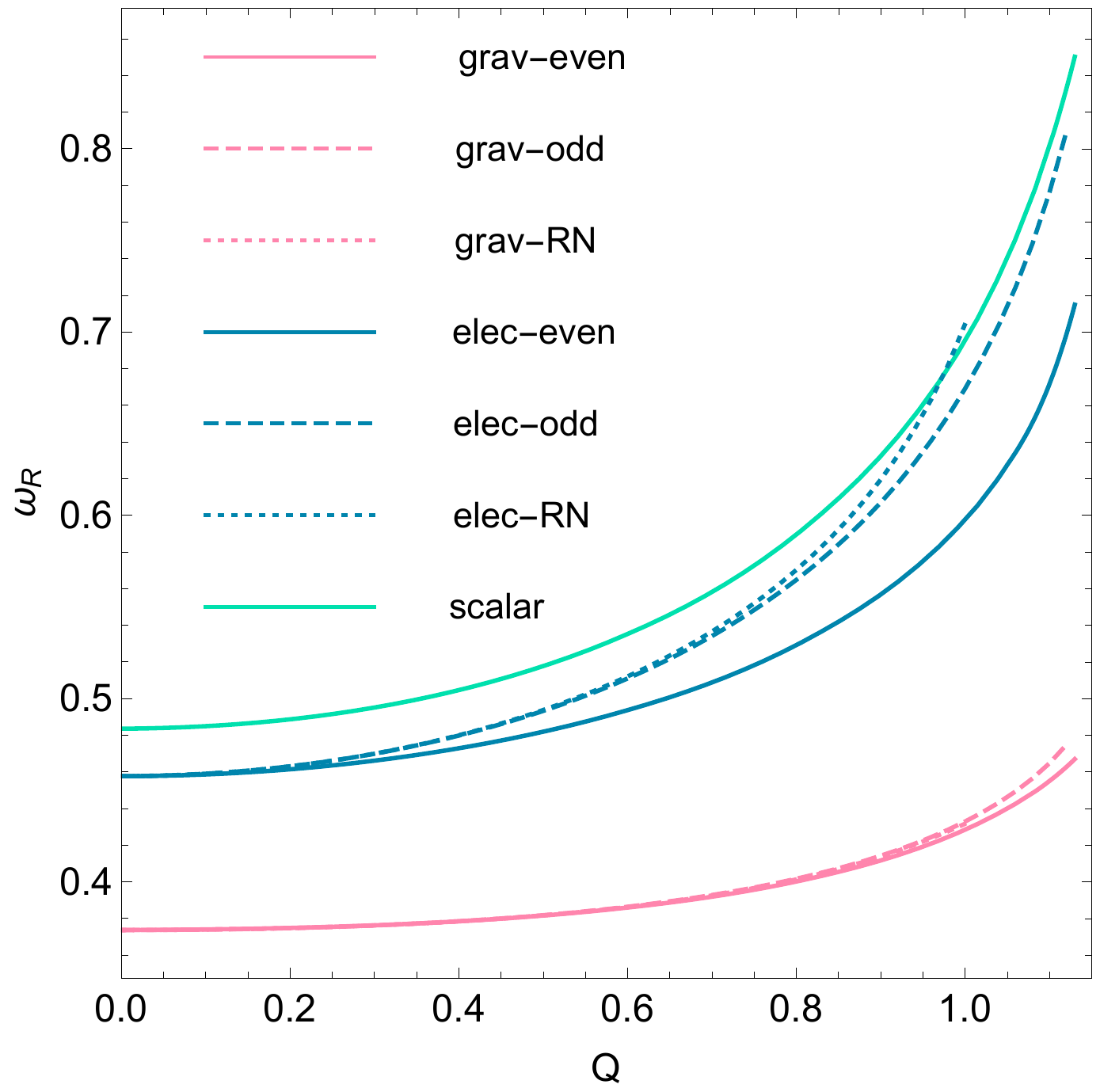}~~~~
\includegraphics[scale=.5]{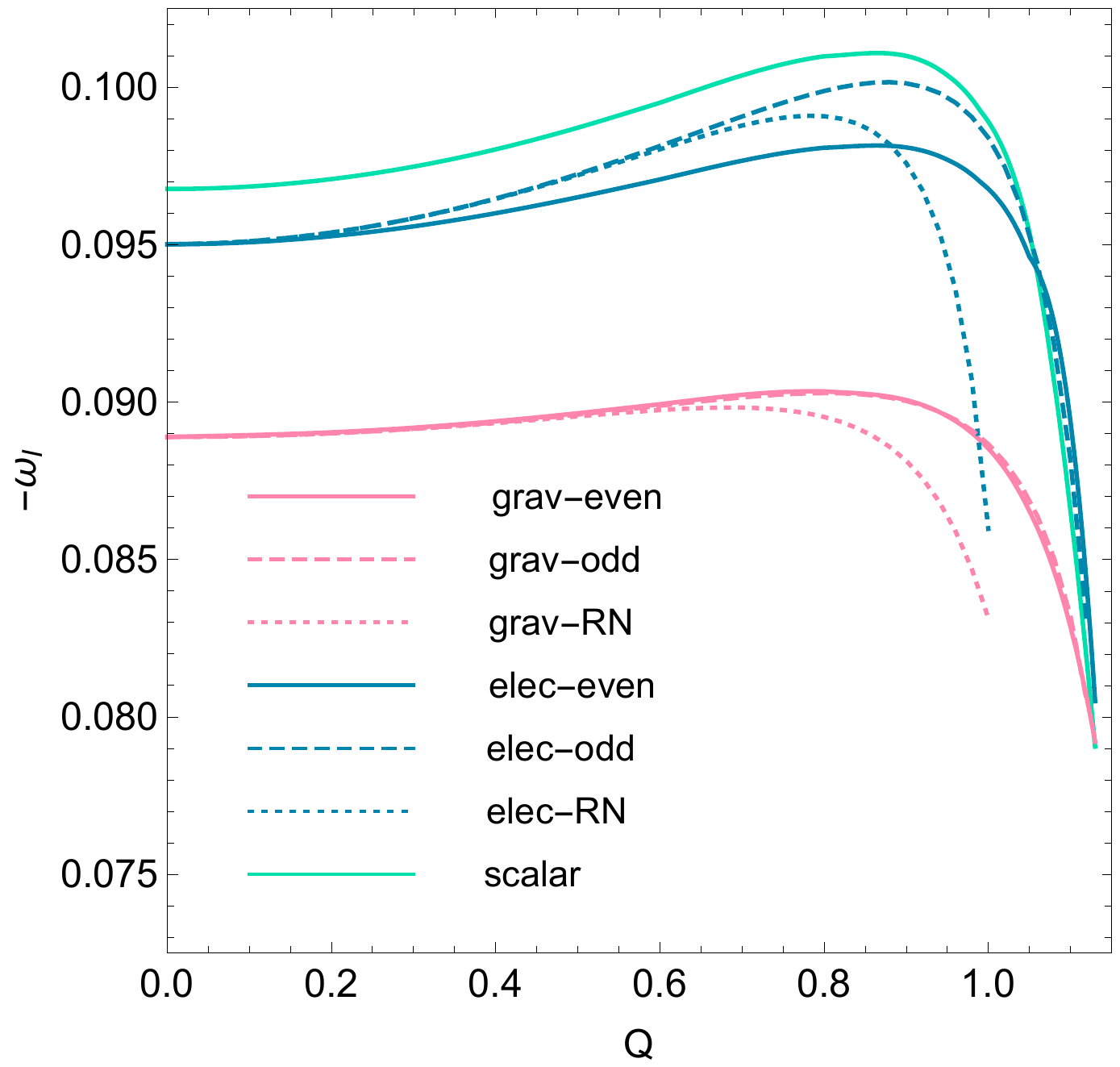}
\caption{Fundamental mode (lowest QNMs), for $\ell=2$ and $a=0.5$, as a function of the charge $Q$, in units $M=1$. In pink, we have the modes associated to the gravitational sector, which reduce to the Regge-Wheeler and Zerilli potentials. In blue, the QNM associated to the electromagnetic sector and finally in blue the mode associated to the scalar field. These perturbations are compared to the Reissner-Nordstrom black hole in dotted line. The QNM have been calculated with a step size of $0.0025$ for the charge. }
\label{QNM-a05}
\end{figure}

Now, we turn to the problem of even-parity perturbations. 
The no-ghost condition (\ref{Stability-Condition}) is trivially satisfied
\begin{equation}
    2 P_1 - f_1 = \frac{2 r_-^2}{(2 r - r_-)^2} > 0 \, .
\end{equation}
Therefore, even-parity perturbations are well defined. It can also be check that all perturbations propagate at the speed of light. In order to study completely the stability of these modes, we need to compute the matrix $S$ (\ref{eq:matrixS}) in order to obtain the potential associated to these perturbations. Unfortunately, in these type of models, the integration of eq. (\ref{eq:matrixS}) is not easy\footnote{The matrices obtained are extremely large and therefore not very useful.} and therefore we will rely to a numerical analysis. For that, we integrate numerically eq.(\ref{eq:matrixS}) assuming any initial conditions as soon as $\det(S)\neq 0$. From which we can easily obtain the potential (\ref{eq:Vscalar}). We found that the eigenvectors of the matrix (\ref{eq:Vscalar}) $-\mathbf{S}^{-1}\mathbf{K}^{-1}\mathbf{B}\mathbf{S}$ are constants and therefore the system has the same eigenvalues than the matrix $-\mathbf{K}^{-1}\mathbf{B}$. As a consequence, we do not need to calculate the matrix $S$, but only the eigenvalues of the matrix $-\mathbf{K}^{-1}\mathbf{B}$. Each eigenvalues will be the effective potential of one of the three type of perturbations. From these effective potentials, we have calculated the QNMs using the sixth order WKB method \cite{Schutz:1985km,Iyer:1986np,Konoplya:2003ii,Konoplya:2011qq}. We see in the Fig.(\ref{QNM-GHS}), the real and imaginary part of the QNMs associated to this black hole compared to the Reissner-Nordstr\"om solution. We see that contrary to the latter, the electric charge breaks the isospectrality\footnote{Same spectrum of quasinormal modes for even and odd perturbations.}. From the numerical results, we found that $\text{Im}(\omega)<0$, for a large number of $\ell$, and therefore we conclude that the black hole is linearly stable. As an example, we have represented, in Figs.(\ref{QNM-GHS},\ref{QNM-a05}), the real and imaginary parts of the QNM for $a=1$ and $a=0.5$ respectively. These results coincide with \cite{Ferrari01} for $a=1$ and with \cite{Brito:2018hjh,Blazquez-Salcedo:2019nwd} in the case $a=0.5$

\section{Conclusions and discussion}

We have studied the stability of static spherically symmetric black holes in generalized Einstein-Maxwell-scalar theories with an electrically charged Maxwell field. We have obtained the master equations in the odd and even parity sectors. We found that the ghost free conditions reduce to 
\begin{align}
    f_1>0\, , \qquad f_{2,F}>0\, , \qquad 3 {f_1}^2_{,\phi}+2 f_1{f_2}_{,X} >0  \, .
\end{align}
We found that four degrees of freedom propagate at the speed of light, while a degree of freedom associated to the scalar field could propagate faster or slower than the speed of light, its expression is given in eq.(\ref{eq:speed}), from which we obtained the subclass of generalized Einstein-Maxwell-scalar theories where all degrees of freedom propagate at the speed of light eq. (\ref{eq:speed1}). Imposing the ghost free conditions, we found that odd-parity perturbations are always stable and we obtained explicitly the effective potential matrix. Considering the even-parity sector, we derived the master equation and the procedure to calculate the effective potential matrix for a given model. We also found that assuming ghost free conditions, the perturbations are unstable if $f_{2,F}+2\bar{F}f_{2,FF}<0$ for models with ${f_2}_{,XF}=0$. Finally, we have applied this formalism for a large number of BH for which we could easily obtain the stability conditions and calculate the QNMs.

We could recover all previously derived results in the literature which guarantees the correctness of our calculations. As an application, the paper can be used to study the QNMs of gravitational waves, electromagnetic radiation and scalar radiation. The presence of these three fields appears naturally in higher dimensional theories of gravity such as supergravity or string theory. Interestingly, we didn't find any stable hairy black hole in scalar tensor theories.  

In order to facilitate future analysis of black hole stability in generalized Einstein-Maxwell-dilaton gravity, all the perturbed equations are listed in a {\scshape Mathematica}\textsuperscript{\textregistered} notebook available online \cite{mathematica}.

\newpage

\appendix
\section{Even-parity coefficients\label{EvenL-coe}}

The coefficients which appear in the second-order Lagrangian~(\ref{L-evenperturbation})
are

\small
\begin{minipage}{0.45\textwidth}
\begin{align*}
a_1 & = -\frac{C}{4}\sqrt{\frac{A}{B}} \qty(\cF + \cE_A)			\\
a_2 & = -\frac12 \sqrt{AB} \Xi										\\
a_3 & = \sqrt{AB} f_1												\\
a_4 & = a_2' + \frac{C}{2} \sqrt{\frac{A}{B}} (\cF + \cM - \cE_B)	\\
a_5 & = -\frac{1}{2}\sqrt{\frac{A}{B}} f_1							\\
a_6 & = a_3' + \qty(\frac{C'}{2C} - \frac{A'}{2A}) a_3 				\\
a_7 & = -C \sqrt{\frac{A}{B}} \frac{\partial \cE_A}{\partial\phi} \\ 
a_8 & = -\sqrt{\frac{A}{B}} \frac{\mathcal{Z}}{C}					\\
a_9 & = \frac{\sqrt{A} C}{\mathcal{Z}}\qty[\frac{\sqrt{B}}{C} \mathcal{Z}^2]' + \cJ - \sqrt{\frac{A}{B}}\frac{C}{\phi'} \cM		\\
a_{10} & = \sqrt{AB} \mathcal{Z}									\\
a_{11} & = -\sqrt{\frac{A}{B}}\frac{C}{\bar{A}_0'} \cF					\\
b_1 & = \sqrt{\frac{B}{A}} C \cE_B				\\
b_2 & = \frac12 \sqrt{\frac{B}{A}} f_1				\\
b_3 & = \sqrt{\frac{B}{A}} \Xi				\\
b_4 & = -2 b_2				\\
b_5 & = \frac{2}{A}\qty(a_{10}' - a_9) - \frac{2 C \cM}{\sqrt{AB} \phi'}	\\
b_6 & = -2\sqrt{\frac{B}{A}} \mathcal{Z} 			\\
c_1 &= \frac{C}{4}\sqrt{\frac{A}{B}} \left[ 2\Sigma + 2B \qty(\frac{A'}{2A} + \frac{C'}{C})f_1' - {\cal E}_B  \right. \\ & \qquad \left. + \qty( \frac{A'}{A} + \frac{C'}{2C} )\frac{B C'}{C} f_1 - \cF - 2\cM \right]	
\end{align*}
\end{minipage} \hfill
\begin{minipage}{0.45\textwidth}
\begin{align*}
c_2 & = \frac{\sqrt{AB}}{C} \qty( f_1 - C \cE_B) 									\\
c_3 & = -\frac12 \sqrt{AB}\qty[ f_1 \qty(\frac{A'}{A} + \frac{C'}{C}) + 2f_1' ]		\\
c_4 & = \frac12 \sqrt{\frac{B}{A}} f_1 												\\
c_5 & = C\sqrt{\frac{A}{B}} \frac{\partial \cE_B}{\partial \phi}					\\
c_6 & = -a_8					\\
c_7 & = -\sqrt{AB}\qty[\qty(\frac{A'}{2A} + \frac{C'}{C}) \mathcal{Z} + \frac{C}{B\phi'} (2 \Sigma - \cM)]	\\
c_8 & = -\frac{\mathcal{Z}}{\sqrt{AB}}												\\
c_9 & = \frac{2 \cJ}{C} + 2\sqrt{ABC} \qty(\frac{\mathcal{Z}}{C^{3/2}})'    		\\
c_{10} & = \frac{2\sqrt{AB}}{C} \mathcal{Z}											\\
c_{11} & = \sqrt{\frac{A}{B}} \frac{C}{\bar{A}_0'} \qty( \cF  + \cM)						\\
c_{12} & = -\frac{2 Q}{C} 														\\
d_1 & = -\sqrt{\frac{A}{B}} \frac{C}{\bar{A}_0'^2} \cF	\\
d_2 & = \frac{Q}{BC \bar{A}_0'}							\\
d_3 & = - AB \, d_2									\\
e_1 & = \frac{\cJ}{AB \phi'}					\\
e_2 & = -\sqrt{AB} C \frac{\Sigma}{\bar{X}}		\\
e_3 & = \frac{\p \cE_\phi}{\p \phi }			\\
e_4 & = -\frac{\cJ}{B C \phi'}					\\
e_5 & = 2C \bar{A}_0'\sqrt{\frac{B}{A}} f_{2, \phi F}	\\
e_6 & = -\frac{2 C}{\bar{A}_0' \phi'}\sqrt{\frac{A}{B}} \cM
\end{align*}
\end{minipage}

\normalsize
where we have defined
\begin{align*}
\Sigma &= \bar{X} \qty(  {f_2}_X + 2 \bar{X} {f_2}_{XX})	\, , \qquad \quad
\Xi = (C f_1)' \\
\cM &= -4 \bar{X} \bar{F} f_{2,FX}  						\, , \qquad\qquad\qquad
\cF = -2 \bar{F} \qty(  {f_2}_F + 2 \bar{F} {f_2}_{FF})		\, , \qquad 
\mathcal{Z} = C f_{1,\phi}\, , \qquad.
\end{align*}

\section{Energy conditions}
\label{energy}
Considering the generalized Einstein-Maxwell-dilaton action (\ref{action}), we obtain the GR-like equation $G_{\mu\nu}=T_{\mu\nu}$ with
\begin{align}
    f_1 T_{\mu\nu}=f_2 g_{\mu\nu}-g_{\mu\nu} \Box f_1+\nabla_{\mu\nu}f_1+\phi_{,\mu}\phi_{,\nu}f_{2,X}+F_{\mu\alpha}F_\nu^{~\alpha} f_{2,F} \, .
\end{align}
Because the energy momentum tensor is diagonal we can interpret these components as the energy density and the three principal pressures in that frame in which the energy-momentum tensor is diagonal. We have
\begin{align}
    T_0^{~0}=-\rho\,,\qquad T_1^{~1}=P_r\,,\qquad T_2^{~2}=T_3^{~3}=P_t  \, ,
\end{align}
where $\rho$ is the energy density, $P_r$ is the radial pressure and $P_t$ is the tangential pressure. 

The energy conditions can be geometric if related directly to the curvature tensor and physical if related to the fluid energy-momentum tensor. In our case, because we will assume that the additional fields can be interpreted as an additional fluid, the equation is reduced to a GR type equation and therefore the geometric and the physical energy conditions are equivalent and described by the following relations
\\
\newline
\textbf{Null energy condition (NEC)}
\begin{align}
    \rho+P_r \geq 0\,, \qquad \rho+P_t \geq 0
\end{align}
\textbf{Weak energy condition (WEC)}
\begin{align}
     \rho \geq 0\,, \qquad \rho+P_r \geq 0\,, \qquad \rho+P_t \geq 0
\end{align}
\textbf{Strong energy condition (SEC)}
\begin{align}
     \rho +P_r+2P_t\geq 0\,, \qquad \rho+P_r \geq 0\,, \qquad \rho+P_t \geq 0
\end{align}
\textbf{Dominant energy condition (DEC)}
\begin{align}
     \rho \geq 0\,, \qquad |P_r| \leq \rho\,, \qquad |P_t| \leq \rho
\end{align}
which can be easily written for generalized Einstein-Maxwell-dilaton knowing that
\begin{align}
&   f_1 \rho=-f_2+2Ff_{2,F}+\frac{\sqrt{B}}{C}\Bigl(\sqrt{B}Cf_1'\Bigr)' \, ,\\
&   f_1 P_r=f_2-2Ff_{2,F}-2Xf_{2,X}-\frac{B(AC^2)'}{2AC^2}f_1' \, ,\\
&   f_1 P_t=f_2-\sqrt{\frac{B}{AC}}\Bigl(\sqrt{ABC}f_1'\Bigr)' \, .
\end{align}
In the simplest case of a minimal coupling, i.e. $f_1$ constant and positive. We have
\\
\newline
\textbf{Null energy condition (NEC)}
\begin{align}
    f_{2,X} \geq 0\,, \qquad f_{2,F} \geq 0
\end{align}
\textbf{Weak energy condition (WEC)}
\begin{align}
     2Ff_{2,F}-f_2 \geq 0\,, \qquad f_{2,X} \geq 0\,, \qquad f_{2,F} \geq 0
\end{align}
\textbf{Strong energy condition (SEC)}
\begin{align}
     f_2-Xf_{2,X}\geq 0\,, \qquad f_{2,X} \geq 0\,, \qquad f_{2,F} \geq 0
\end{align}
\textbf{Dominant energy condition (DEC)}
\begin{align}
     2Ff_{2,F}-f_2 \geq 0\,, \qquad |f_2-2Ff_{2,F}-2Xf_{2,X}| \leq 2Ff_{2,F}-f_2\,, \qquad |f_2| \leq 2Ff_{2,F}-f_2
\end{align}


\begin{thebibliography}{10}

\bibitem{Regge57}
T.~Regge and J.A.~Wheeler, \emph{Stability of a schwarzschild singularity},
  \href{https://doi.org/10.1103/PhysRev.108.1063}{\emph{Phys. Rev.} {\bfseries
  108} (1957) 1063}.

\bibitem{Zerilli70}
F.J.~Zerilli, \emph{{Effective potential for even parity Regge-Wheeler
  gravitational perturbation equations}},
  \href{https://doi.org/10.1103/PhysRevLett.24.737}{\emph{Phys. Rev. Lett.}
  {\bfseries 24} (1970) 737}.

\bibitem{Moncrief74-a}
V.~Moncrief, \emph{Odd-parity stability of a reissner-nordstr\"om black hole},
  \href{https://doi.org/10.1103/PhysRevD.9.2707}{\emph{Phys. Rev. D} {\bfseries
  9} (1974) 2707}.

\bibitem{Moncrief74-b}
V.~Moncrief, \emph{Stability of reissner-nordstr\"om black holes},
  \href{https://doi.org/10.1103/PhysRevD.10.1057}{\emph{Phys. Rev. D}
  {\bfseries 10} (1974) 1057}.

\bibitem{Teukolsky:1972my}
S.A.~Teukolsky, \emph{{Rotating black holes - separable wave equations for
  gravitational and electromagnetic perturbations}},
  \href{https://doi.org/10.1103/PhysRevLett.29.1114}{\emph{Phys. Rev. Lett.}
  {\bfseries 29} (1972) 1114}.

\bibitem{Teukolsky:1973ha}
S.A.~Teukolsky, \emph{{Perturbations of a rotating black hole. 1. Fundamental
  equations for gravitational electromagnetic and neutrino field
  perturbations}}, \href{https://doi.org/10.1086/152444}{\emph{Astrophys. J.}
  {\bfseries 185} (1973) 635}.

\bibitem{DeFelice:2011ka}
A.~De~Felice, T.~Suyama and T.~Tanaka, \emph{{Stability of Schwarzschild-like
  solutions in f(R,G) gravity models}},
  \href{https://doi.org/10.1103/PhysRevD.83.104035}{\emph{Phys. Rev. D}
  {\bfseries 83} (2011) 104035}
  [\href{https://arxiv.org/abs/1102.1521}{{\ttfamily 1102.1521}}].

\bibitem{Takahashi:2010ye}
T.~Takahashi and J.~Soda, \emph{{Master Equations for Gravitational
  Perturbations of Static Lovelock Black Holes in Higher Dimensions}},
  \href{https://doi.org/10.1143/PTP.124.911}{\emph{Prog. Theor. Phys.}
  {\bfseries 124} (2010) 911}
  [\href{https://arxiv.org/abs/1008.1385}{{\ttfamily 1008.1385}}].

\bibitem{Ganguly:2017ort}
A.~Ganguly, R.~Gannouji, M.~Gonzalez-Espinoza and C.~Pizarro-Moya, \emph{{Black
  hole stability under odd-parity perturbations in Horndeski gravity}},
  \href{https://doi.org/10.1088/1361-6382/aac8a0}{\emph{Class. Quant. Grav.}
  {\bfseries 35} (2018) 145008}
  [\href{https://arxiv.org/abs/1710.07669}{{\ttfamily 1710.07669}}].

\bibitem{Gleyzes:2014dya}
J.~Gleyzes, D.~Langlois, F.~Piazza and F.~Vernizzi, \emph{{Healthy theories
  beyond Horndeski}},
  \href{https://doi.org/10.1103/PhysRevLett.114.211101}{\emph{Phys. Rev. Lett.}
  {\bfseries 114} (2015) 211101}
  [\href{https://arxiv.org/abs/1404.6495}{{\ttfamily 1404.6495}}].

\bibitem{Gleyzes:2014qga}
J.~Gleyzes, D.~Langlois, F.~Piazza and F.~Vernizzi, \emph{{Exploring
  gravitational theories beyond Horndeski}},
  \href{https://doi.org/10.1088/1475-7516/2015/02/018}{\emph{JCAP} {\bfseries
  02} (2015) 018} [\href{https://arxiv.org/abs/1408.1952}{{\ttfamily
  1408.1952}}].

\bibitem{Langlois:2015cwa}
D.~Langlois and K.~Noui, \emph{{Degenerate higher derivative theories beyond
  Horndeski: evading the Ostrogradski instability}},
  \href{https://doi.org/10.1088/1475-7516/2016/02/034}{\emph{JCAP} {\bfseries
  02} (2016) 034} [\href{https://arxiv.org/abs/1510.06930}{{\ttfamily
  1510.06930}}].

\bibitem{Papallo:2017ddx}
G.~Papallo, \emph{{On the hyperbolicity of the most general Horndeski theory}},
  \href{https://doi.org/10.1103/PhysRevD.96.124036}{\emph{Phys. Rev. D}
  {\bfseries 96} (2017) 124036}
  [\href{https://arxiv.org/abs/1710.10155}{{\ttfamily 1710.10155}}].

\bibitem{Papallo:2017qvl}
G.~Papallo and H.S.~Reall, \emph{{On the local well-posedness of Lovelock and
  Horndeski theories}},
  \href{https://doi.org/10.1103/PhysRevD.96.044019}{\emph{Phys. Rev. D}
  {\bfseries 96} (2017) 044019}
  [\href{https://arxiv.org/abs/1705.04370}{{\ttfamily 1705.04370}}].

\bibitem{Kovacs:2020ywu}
A.D.~Kov\'acs and H.S.~Reall, \emph{{Well-posed formulation of Lovelock and
  Horndeski theories}},
  \href{https://doi.org/10.1103/PhysRevD.101.124003}{\emph{Phys. Rev. D}
  {\bfseries 101} (2020) 124003}
  [\href{https://arxiv.org/abs/2003.08398}{{\ttfamily 2003.08398}}].

\bibitem{Gibbons88}
G.~Gibbons and K.~ichi Maeda, \emph{Black holes and membranes in
  higher-dimensional theories with dilaton fields},
  \href{https://doi.org/https://doi.org/10.1016/0550-3213(88)90006-5}{\emph{Nuclear
  Physics B} {\bfseries 298} (1988) 741}.

\bibitem{Garfinkle91}
D.~Garfinkle, G.T.~Horowitz and A.~Strominger, \emph{Charged black holes in
  string theory}, \href{https://doi.org/10.1103/PhysRevD.43.3140}{\emph{Phys.
  Rev. D} {\bfseries 43} (1991) 3140}.

\bibitem{mathematica}
\url{https://github.com/YolbeikerRB/Even-parity_Perturbations.git}.

\bibitem{DeFelice2016}
A.~De~Felice, L.~Heisenberg, R.~Kase, S.~Tsujikawa, Y.-l.~Zhang and G.-B.~Zhao,
  \emph{{Screening fifth forces in generalized Proca theories}},
  \href{https://doi.org/10.1103/PhysRevD.93.104016}{\emph{Phys. Rev.}
  {\bfseries D93} (2016) 104016}
  [\href{https://arxiv.org/abs/1602.00371}{{\ttfamily 1602.00371}}].

\bibitem{Gibbons:2002pq}
G.~Gibbons and S.A.~Hartnoll, \emph{{A Gravitational instability in higher
  dimensions}}, \href{https://doi.org/10.1103/PhysRevD.66.064024}{\emph{Phys.
  Rev. D} {\bfseries 66} (2002) 064024}
  [\href{https://arxiv.org/abs/hep-th/0206202}{{\ttfamily hep-th/0206202}}].

\bibitem{Motohashi11}
H.~Motohashi and T.~Suyama, \emph{Black hole perturbation in parity violating
  gravitational theories},
  \href{https://doi.org/10.1103/physrevd.84.084041}{\emph{Physical Review D}
  {\bfseries 84} (2011) }.

\bibitem{Chandrasekhar98}
S.~Chandrasekhar, \emph{The Mathematical Theory of Black Holes}, International
  series of monographs on physics, Clarendon Press (1998).

\bibitem{DeFelice11}
A.~De~Felice, T.~Suyama and T.~Tanaka, \emph{Stability of schwarzschild-like
  solutions inf(r,g)gravity models},
  \href{https://doi.org/10.1103/physrevd.83.104035}{\emph{Physical Review D}
  {\bfseries 83} (2011) } [\href{https://arxiv.org/abs/1102.1521}{{\ttfamily
  1102.1521}}].

\bibitem{Ganguly18}
A.~Ganguly, R.~Gannouji, M.~Gonzalez-Espinoza and C.~Pizarro-Moya, \emph{Black
  hole stability under odd-parity perturbations in horndeski gravity},
  \href{https://doi.org/10.1088/1361-6382/aac8a0}{\emph{Classical and Quantum
  Gravity} {\bfseries 35} (2018) 145008}
  [\href{https://arxiv.org/abs/1710.07669}{{\ttfamily 1710.07669}}].

\bibitem{wald1}
R.~Wald, \emph{{Note on the stability of the schwarzschild metric}},
  \href{https://doi.org/10.1063/1.524181}{\emph{J. Math. Phys.} {\bfseries 20}
  (1979) 1056}.

\bibitem{wald2}
R.~Wald, \emph{{Erratum: Note on the stability of the schwarzschild metric}},
  \href{https://doi.org/10.1063/1.524324}{\emph{J. Math. Phys.} {\bfseries 21}
  (1980) 218}.

\bibitem{Takahashi:2010gz}
T.~Takahashi and J.~Soda, \emph{{Catastrophic Instability of Small Lovelock
  Black Holes}}, \href{https://doi.org/10.1143/PTP.124.711}{\emph{Prog. Theor.
  Phys.} {\bfseries 124} (2010) 711}
  [\href{https://arxiv.org/abs/1008.1618}{{\ttfamily 1008.1618}}].

\bibitem{Moreno:2002gg}
C.~Moreno and O.~Sarbach, \emph{{Stability properties of black holes in
  selfgravitating nonlinear electrodynamics}},
  \href{https://doi.org/10.1103/PhysRevD.67.024028}{\emph{Phys. Rev. D}
  {\bfseries 67} (2003) 024028}
  [\href{https://arxiv.org/abs/gr-qc/0208090}{{\ttfamily gr-qc/0208090}}].

\bibitem{Kobayashi2014}
T.~Kobayashi, H.~Motohashi and T.~Suyama, \emph{{Black hole perturbation in the
  most general scalar-tensor theory with second-order field equations II: the
  even-parity sector}},
  \href{https://doi.org/10.1103/PhysRevD.89.084042}{\emph{Phys. Rev. D}
  {\bfseries 89} (2014) 084042}
  [\href{https://arxiv.org/abs/1402.6740}{{\ttfamily 1402.6740}}].

\bibitem{Afshordi:2006ad}
N.~Afshordi, D.J.H.~Chung and G.~Geshnizjani, \emph{{Cuscuton: A Causal Field
  Theory with an Infinite Speed of Sound}},
  \href{https://doi.org/10.1103/PhysRevD.75.083513}{\emph{Phys. Rev. D}
  {\bfseries 75} (2007) 083513}
  [\href{https://arxiv.org/abs/hep-th/0609150}{{\ttfamily hep-th/0609150}}].

\bibitem{Adams:2006sv}
A.~Adams, N.~Arkani-Hamed, S.~Dubovsky, A.~Nicolis and R.~Rattazzi,
  \emph{{Causality, analyticity and an IR obstruction to UV completion}},
  \href{https://doi.org/10.1088/1126-6708/2006/10/014}{\emph{JHEP} {\bfseries
  10} (2006) 014} [\href{https://arxiv.org/abs/hep-th/0602178}{{\ttfamily
  hep-th/0602178}}].

\bibitem{Nicolis:2009qm}
A.~Nicolis, R.~Rattazzi and E.~Trincherini, \emph{{Energy's and amplitudes'
  positivity}}, \href{https://doi.org/10.1007/JHEP05(2010)095}{\emph{JHEP}
  {\bfseries 05} (2010) 095} [\href{https://arxiv.org/abs/0912.4258}{{\ttfamily
  0912.4258}}].

\bibitem{Melville:2019wyy}
S.~Melville and J.~Noller, \emph{{Positivity in the Sky: Constraining dark
  energy and modified gravity from the UV}},
  \href{https://doi.org/10.1103/PhysRevD.101.021502}{\emph{Phys. Rev. D}
  {\bfseries 101} (2020) 021502}
  [\href{https://arxiv.org/abs/1904.05874}{{\ttfamily 1904.05874}}].

\bibitem{Volkov:1994dq}
M.S.~Volkov and D.V.~Galtsov, \emph{{Odd parity negative modes of Einstein
  Yang-Mills black holes and sphalerons}},
  \href{https://doi.org/10.1016/0370-2693(95)80005-I}{\emph{Phys. Lett. B}
  {\bfseries 341} (1995) 279}
  [\href{https://arxiv.org/abs/hep-th/9409041}{{\ttfamily hep-th/9409041}}].

\bibitem{Bardeen}
J.M.~Bardeen, \emph{Non-singular general-relativistic gravitational collapse},
  in \emph{Proc. Int. Conf. GR5, Tbilisi}, vol.~174, 1968.

\bibitem{Ayon-Beato:2000mjt}
E.~Ayon-Beato and A.~Garcia, \emph{{The Bardeen model as a nonlinear magnetic
  monopole}}, \href{https://doi.org/10.1016/S0370-2693(00)01125-4}{\emph{Phys.
  Lett. B} {\bfseries 493} (2000) 149}
  [\href{https://arxiv.org/abs/gr-qc/0009077}{{\ttfamily gr-qc/0009077}}].

\bibitem{Salazar:1987ap}
I.H.~Salazar, A.~Garcia and J.~Plebanski, \emph{{Duality Rotations and Type $D$
  Solutions to Einstein Equations With Nonlinear Electromagnetic Sources}},
  \href{https://doi.org/10.1063/1.527430}{\emph{J. Math. Phys.} {\bfseries 28}
  (1987) 2171}.

\bibitem{Bronnikov:2000vy}
K.A.~Bronnikov, \emph{{Regular magnetic black holes and monopoles from
  nonlinear electrodynamics}},
  \href{https://doi.org/10.1103/PhysRevD.63.044005}{\emph{Phys. Rev. D}
  {\bfseries 63} (2001) 044005}
  [\href{https://arxiv.org/abs/gr-qc/0006014}{{\ttfamily gr-qc/0006014}}].

\bibitem{Duff:1974ud}
M.J.~Duff, \emph{{Quantum corrections to the schwarzschild solution}},
  \href{https://doi.org/10.1103/PhysRevD.9.1837}{\emph{Phys. Rev. D} {\bfseries
  9} (1974) 1837}.

\bibitem{Bjerrum-Bohr:2002fji}
N.E.J.~Bjerrum-Bohr, J.F.~Donoghue and B.R.~Holstein, \emph{{Quantum
  corrections to the Schwarzschild and Kerr metrics}},
  \href{https://doi.org/10.1103/PhysRevD.68.084005}{\emph{Phys. Rev. D}
  {\bfseries 68} (2003) 084005}
  [\href{https://arxiv.org/abs/hep-th/0211071}{{\ttfamily hep-th/0211071}}].

\bibitem{Hayward:2005gi}
S.A.~Hayward, \emph{{Formation and evaporation of regular black holes}},
  \href{https://doi.org/10.1103/PhysRevLett.96.031103}{\emph{Phys. Rev. Lett.}
  {\bfseries 96} (2006) 031103}
  [\href{https://arxiv.org/abs/gr-qc/0506126}{{\ttfamily gr-qc/0506126}}].

\bibitem{Frolov:2016pav}
V.P.~Frolov, \emph{{Notes on nonsingular models of black holes}},
  \href{https://doi.org/10.1103/PhysRevD.94.104056}{\emph{Phys. Rev. D}
  {\bfseries 94} (2016) 104056}
  [\href{https://arxiv.org/abs/1609.01758}{{\ttfamily 1609.01758}}].

\bibitem{Tseytlin:1999dj}
A.A.~Tseytlin, \emph{{Born-Infeld action, supersymmetry and string theory}},
  \href{https://doi.org/10.1142/9789812793850_0025}{\emph{The Many Faces of the
  Superworld} (1999) 417}
  [\href{https://arxiv.org/abs/hep-th/9908105}{{\ttfamily hep-th/9908105}}].

\bibitem{Cederwall:1996uu}
M.~Cederwall, A.~von Gussich, A.R.~Mikovic, B.E.W.~Nilsson and A.~Westerberg,
  \emph{{On the Dirac-Born-Infeld action for d-branes}},
  \href{https://doi.org/10.1016/S0370-2693(96)01367-6}{\emph{Phys. Lett. B}
  {\bfseries 390} (1997) 148}
  [\href{https://arxiv.org/abs/hep-th/9606173}{{\ttfamily hep-th/9606173}}].

\bibitem{Fernando:2004pc}
S.~Fernando, \emph{{Gravitational perturbation and quasi-normal modes of
  charged black holes in Einstein-Born-Infeld gravity}},
  \href{https://doi.org/10.1007/s10714-005-0044-9}{\emph{Gen. Rel. Grav.}
  {\bfseries 37} (2005) 585}
  [\href{https://arxiv.org/abs/hep-th/0407062}{{\ttfamily hep-th/0407062}}].

\bibitem{Fernando:2005bc}
S.~Fernando and C.~Holbrook, \emph{{Stability and quasi normal modes of charged
  black holes in Born-Infeld gravity}},
  \href{https://doi.org/10.1007/s10773-005-9024-9}{\emph{Int. J. Theor. Phys.}
  {\bfseries 45} (2006) 1630}
  [\href{https://arxiv.org/abs/hep-th/0501138}{{\ttfamily hep-th/0501138}}].

\bibitem{Armendariz-Picon:1999hyi}
C.~Armendariz-Picon, T.~Damour and V.F.~Mukhanov, \emph{{k - inflation}},
  \href{https://doi.org/10.1016/S0370-2693(99)00603-6}{\emph{Phys. Lett. B}
  {\bfseries 458} (1999) 209}
  [\href{https://arxiv.org/abs/hep-th/9904075}{{\ttfamily hep-th/9904075}}].

\bibitem{Armendariz-Picon:2000nqq}
C.~Armendariz-Picon, V.F.~Mukhanov and P.J.~Steinhardt, \emph{{A Dynamical
  solution to the problem of a small cosmological constant and late time cosmic
  acceleration}},
  \href{https://doi.org/10.1103/PhysRevLett.85.4438}{\emph{Phys. Rev. Lett.}
  {\bfseries 85} (2000) 4438}
  [\href{https://arxiv.org/abs/astro-ph/0004134}{{\ttfamily
  astro-ph/0004134}}].

\bibitem{Ravndal:2004ym}
F.~Ravndal, \emph{{Scalar gravitation and extra dimensions}}, {\emph{Comment.
  Phys. Math. Soc. Sci. Fenn.} {\bfseries 166} (2004) 151}
  [\href{https://arxiv.org/abs/gr-qc/0405030}{{\ttfamily gr-qc/0405030}}].

\bibitem{Green:1987sp}
M.B.~Green, J.H.~Schwarz and E.~Witten, \emph{{SUPERSTRING THEORY. VOL. 1:
  INTRODUCTION}}, Cambridge Monographs on Mathematical Physics, Cambridge
  University Press (7, 1988).

\bibitem{Garousi:2000tr}
M.R.~Garousi, \emph{{Tachyon couplings on nonBPS D-branes and Dirac-Born-Infeld
  action}}, \href{https://doi.org/10.1016/S0550-3213(00)00361-8}{\emph{Nucl.
  Phys. B} {\bfseries 584} (2000) 284}
  [\href{https://arxiv.org/abs/hep-th/0003122}{{\ttfamily hep-th/0003122}}].

\bibitem{Sen:2002in}
A.~Sen, \emph{{Tachyon matter}},
  \href{https://doi.org/10.1088/1126-6708/2002/07/065}{\emph{JHEP} {\bfseries
  07} (2002) 065} [\href{https://arxiv.org/abs/hep-th/0203265}{{\ttfamily
  hep-th/0203265}}].

\bibitem{Arkani-Hamed:2003pdi}
N.~Arkani-Hamed, H.-C.~Cheng, M.A.~Luty and S.~Mukohyama, \emph{{Ghost
  condensation and a consistent infrared modification of gravity}},
  \href{https://doi.org/10.1088/1126-6708/2004/05/074}{\emph{JHEP} {\bfseries
  05} (2004) 074} [\href{https://arxiv.org/abs/hep-th/0312099}{{\ttfamily
  hep-th/0312099}}].

\bibitem{Arkani-Hamed:2003juy}
N.~Arkani-Hamed, P.~Creminelli, S.~Mukohyama and M.~Zaldarriaga, \emph{{Ghost
  inflation}}, \href{https://doi.org/10.1088/1475-7516/2004/04/001}{\emph{JCAP}
  {\bfseries 04} (2004) 001}
  [\href{https://arxiv.org/abs/hep-th/0312100}{{\ttfamily hep-th/0312100}}].

\bibitem{Will:2014kxa}
C.M.~Will, \emph{{The Confrontation between General Relativity and
  Experiment}}, \href{https://doi.org/10.12942/lrr-2014-4}{\emph{Living Rev.
  Rel.} {\bfseries 17} (2014) 4}
  [\href{https://arxiv.org/abs/1403.7377}{{\ttfamily 1403.7377}}].

\bibitem{Veneziano:1997pz}
G.~Veneziano, \emph{{Inhomogeneous pre - big bang string cosmology}},
  \href{https://doi.org/10.1016/S0370-2693(97)00688-6}{\emph{Phys. Lett. B}
  {\bfseries 406} (1997) 297}
  [\href{https://arxiv.org/abs/hep-th/9703150}{{\ttfamily hep-th/9703150}}].

\bibitem{Hui:2012qt}
L.~Hui and A.~Nicolis, \emph{{No-Hair Theorem for the Galileon}},
  \href{https://doi.org/10.1103/PhysRevLett.110.241104}{\emph{Phys. Rev. Lett.}
  {\bfseries 110} (2013) 241104}
  [\href{https://arxiv.org/abs/1202.1296}{{\ttfamily 1202.1296}}].

\bibitem{Bronnikov:2006qj}
K.A.~Bronnikov, M.S.~Chernakova, J.C.~Fabris, N.~Pinto-Neto and M.E.~Rodrigues,
  \emph{{Cold black holes and conformal continuations}},
  \href{https://doi.org/10.1142/S0218271808011845}{\emph{Int. J. Mod. Phys. D}
  {\bfseries 17} (2008) 25}
  [\href{https://arxiv.org/abs/gr-qc/0609084}{{\ttfamily gr-qc/0609084}}].

\bibitem{Bekenstein:1972ny}
J.D.~Bekenstein, \emph{{Transcendence of the law of baryon-number conservation
  in black hole physics}},
  \href{https://doi.org/10.1103/PhysRevLett.28.452}{\emph{Phys. Rev. Lett.}
  {\bfseries 28} (1972) 452}.

\bibitem{Bekenstein:1995un}
J.D.~Bekenstein, \emph{{Novel
  \textquoteleft{}\textquoteleft{}no-scalar-hair\textquoteright{}\textquoteright{}
  theorem for black holes}},
  \href{https://doi.org/10.1103/PhysRevD.51.R6608}{\emph{Phys. Rev. D}
  {\bfseries 51} (1995) R6608}.

\bibitem{Sudarsky:1995zg}
D.~Sudarsky, \emph{{A Simple proof of a no hair theorem in Einstein Higgs
  theory,}}, \href{https://doi.org/10.1088/0264-9381/12/2/023}{\emph{Class.
  Quant. Grav.} {\bfseries 12} (1995) 579}.

\bibitem{Herdeiro:2015waa}
C.A.R.~Herdeiro and E.~Radu, \emph{{Asymptotically flat black holes with scalar
  hair: a review}}, \href{https://doi.org/10.1142/S0218271815420146}{\emph{Int.
  J. Mod. Phys. D} {\bfseries 24} (2015) 1542014}
  [\href{https://arxiv.org/abs/1504.08209}{{\ttfamily 1504.08209}}].

\bibitem{Bocharova:1970}
N.M.~Bocharova, K.A.~Bronnikov and V.N.~Melnikov, \emph{{An exact solution of
  the system of Einstein equations and mass-free scalar field}}, {\emph{Vestn.
  Mosk. Univ. Ser. III Fiz. Astron.} {\bfseries 6} (1970) 706}.

\bibitem{Bekenstein:1974sf}
J.D.~Bekenstein, \emph{{Exact solutions of Einstein conformal scalar
  equations}}, \href{https://doi.org/10.1016/0003-4916(74)90124-9}{\emph{Annals
  Phys.} {\bfseries 82} (1974) 535}.

\bibitem{Bekenstein:1975ts}
J.D.~Bekenstein, \emph{{Black Holes with Scalar Charge}},
  \href{https://doi.org/10.1016/0003-4916(75)90279-1}{\emph{Annals Phys.}
  {\bfseries 91} (1975) 75}.

\bibitem{Bronnikov:1978mx}
K.A.~Bronnikov and Y.N.~Kireev, \emph{{Instability of Black Holes with Scalar
  Charge}}, \href{https://doi.org/10.1016/0375-9601(78)90030-0}{\emph{Phys.
  Lett. A} {\bfseries 67} (1978) 95}.

\bibitem{Nozawa:2018kfk}
M.~Nozawa, T.~Shiromizu, K.~Izumi and S.~Yamada, \emph{{Divergence equations
  and uniqueness theorem of static black holes}},
  \href{https://doi.org/10.1088/1361-6382/aad206}{\emph{Class. Quant. Grav.}
  {\bfseries 35} (2018) 175009}
  [\href{https://arxiv.org/abs/1805.11385}{{\ttfamily 1805.11385}}].

\bibitem{Ferrari01}
V.~Ferrari, M.~Pauri and F.~Piazza, \emph{Quasinormal modes of charged, dilaton
  black holes},
  \href{https://doi.org/10.1103/physrevd.63.064009}{\emph{Physical Review D}
  {\bfseries 63} (2001) }
  [\href{https://arxiv.org/abs/gr-qc/0005125}{{\ttfamily gr-qc/0005125}}].

\bibitem{Schutz:1985km}
B.F.~Schutz and C.M.~Will, \emph{{Black hole normal modes - A semianalytic
  approach}}, \href{https://doi.org/10.1086/184453}{\emph{Astrophys. J. Lett.}
  {\bfseries 291} (1985) L33}.

\bibitem{Iyer:1986np}
S.~Iyer and C.M.~Will, \emph{{Black Hole Normal Modes: A {WKB} Approach. 1.
  Foundations and Application of a Higher Order {WKB} Analysis of Potential
  Barrier Scattering}},
  \href{https://doi.org/10.1103/PhysRevD.35.3621}{\emph{Phys. Rev. D}
  {\bfseries 35} (1987) 3621}.

\bibitem{Konoplya:2003ii}
R.A.~Konoplya, \emph{{Quasinormal behavior of the d-dimensional Schwarzschild
  black hole and higher order WKB approach}},
  \href{https://doi.org/10.1103/PhysRevD.68.024018}{\emph{Phys. Rev. D}
  {\bfseries 68} (2003) 024018}
  [\href{https://arxiv.org/abs/gr-qc/0303052}{{\ttfamily gr-qc/0303052}}].

\bibitem{Konoplya:2011qq}
R.A.~Konoplya and A.~Zhidenko, \emph{{Quasinormal modes of black holes: From
  astrophysics to string theory}},
  \href{https://doi.org/10.1103/RevModPhys.83.793}{\emph{Rev. Mod. Phys.}
  {\bfseries 83} (2011) 793} [\href{https://arxiv.org/abs/1102.4014}{{\ttfamily
  1102.4014}}].

\bibitem{Brito:2018hjh}
R.~Brito and C.~Pacilio, \emph{{Quasinormal modes of weakly charged
  Einstein-Maxwell-dilaton black holes}},
  \href{https://doi.org/10.1103/PhysRevD.98.104042}{\emph{Phys. Rev. D}
  {\bfseries 98} (2018) 104042}
  [\href{https://arxiv.org/abs/1807.09081}{{\ttfamily 1807.09081}}].

\bibitem{Blazquez-Salcedo:2019nwd}
J.L.~Bl\'azquez-Salcedo, S.~Kahlen and J.~Kunz, \emph{{Quasinormal modes of
  dilatonic Reissner\textendash{}Nordstr\"om black holes}},
  \href{https://doi.org/10.1140/epjc/s10052-019-7535-4}{\emph{Eur. Phys. J. C}
  {\bfseries 79} (2019) 1021}
  [\href{https://arxiv.org/abs/1911.01943}{{\ttfamily 1911.01943}}].

\end{thebibliography}

\providecommand{\href}[2]{#2}\begingroup\raggedright\endgroup
\end{document}